\documentclass[a4paper,fleqn]{cas-sc}

\usepackage[numbers,sort&compress]{natbib}
\setcitestyle{square,numbers}

\def\tsc#1{\csdef{#1}{\textsc{\lowercase{#1}}\xspace}}
\tsc{WGM}
\tsc{QE}
\tsc{EP}
\tsc{PMS}
\tsc{BEC}
\tsc{DE}
\usepackage{bm}

\newproof{pf}{Proof}
\newproof{pot}{Proof of Theorem \ref{thm}}

\begin{document}
\let\WriteBookmarks\relax
\def\floatpagepagefraction{1}
\def\textpagefraction{.001}
\shorttitle{The geometric origin of criticality}
\shortauthors{L. Di Cairano}


\title [mode = title]{The geometric origin of criticality: a universal mechanism in mean-field rotor Hamiltonians}

\author[1]{Loris Di~Cairano}[type=editor,
                        auid=000,bioid=1,
                        orcid=0000-0001-5859-4188]
\cormark[1]
\fnmark[1]
\ead{l.di.cairano.92@gmail.com}


\affiliation[1]{organization={University of Luxembourg, Department of Physics and Materials Science},
                city={ Luxembourg City},
                postcode={L-1511}, 
                country={Luxembourg}}



\begin{abstract}
We introduce a universal criterion for criticality in mean-field rotor Hamiltonians based on the geometric structure of the constant-energy shell. Rather than characterizing the onset of a phase transition through the conventional thermodynamic singularities alone, we show that the relevant information is already encoded in the way the geometry of the shell reorganizes along distinguished collective directions. 

For a broad class of finite-dimensional trigonometric mean-field interactions, the trace of the Weingarten operator (representing the principal curvatures) admits a universal collective expansion in terms of the order-parameter amplitudes. This expansion defines an energy-dependent quadratic form whose eigenmodes identify the geometrically unstable channels of the system. Criticality is then associated with the vanishing of the corresponding curvature coefficients, yielding a direct geometric selection principle for the modes that become unstable at the transition.

In this way, the phase transition in mean-field systems (usually of first- or second-order) is reformulated as a geometric instability phenomenon intrinsic to the microcanonical energy shell. The resulting framework is geometrically universal within the class considered, independent of model-specific details except for a finite set of collective couplings. Moreover, our approach recovers the known critical channels in standard mean-field rotor models while extending naturally to multimode and spectrally coupled cases. These results support a view in which critical behavior can be understood as reorganizations
of energy-shell geometry triggered by a collective restructuring of the underlying energy-shell geometry.

\end{abstract}


\begin{keywords}
Phase transitions\sep
Mean-field rotor Hamiltonians\sep 
Microcanonical ensemble \sep
Energy-shell curvature \sep
Geometric mechanism\sep
Weingarten operator\sep
\sep
\vspace{0.2cm}
\textit{\textbf{Significance}}\sep
Phase transitions are usually described through thermodynamic singularities, but this characterization does not isolate the microscopic origin that triggers criticality. Here we show that, for a broad class of mean-field rotor Hamiltonians, such an origin exists and it is revealed by an universal mechanism. The transition is driven by a change of geometric curvature of the constant-energy. 
\end{keywords}

\maketitle

\tableofcontents

\section{Introduction}

\subsection*{What is the deep origin of a phase transition?}
Phase transitions occupy a singular place in theoretical physics. On the one hand, they are among the collective phenomena best understood within the framework of statistical mechanics: the thermodynamic formalism identifies their thresholds with great effectiveness, describes the different phases, and organizes their macroscopic manifestations in terms of order, fluctuations, and stability. On the other hand, precisely this effectiveness tends to leave in the background a more radical issue, one that no longer concerns only the description of the phenomenon, but its structural status: what truly changes in a system when it crosses a critical threshold?

Within the standard framework, this question is addressed through a set of complementary statistical and thermodynamic theories. In the Landau picture~\cite{Landau1937}, criticality is encoded in the instability and bifurcation structure of an order-parameter functional; in the Lee--Yang theory~\cite{LeeYang1952a,LeeYang1952b}, it is reflected in the distribution of partition-function zeros in the complex plane; and in the renormalization-group framework~\cite{WilsonKogut1974,Wilson1975}, it is understood in terms of scale dependence, universality, and the flow of effective descriptions across scales. Taken together, these approaches support a common view: criticality is an emergent phenomenon associated with a collective reorganization of statistical weights, fluctuations, and dominant sectors of the space of states. This picture is deep and extraordinarily fruitful, but it does not necessarily exhaust the question. Does such a statistical structure constitute the original level of the phenomenon, or does it instead represent the descriptive level at which a deeper reorganization becomes observable? In other words, beyond its thermodynamic identification, is there an underlying structural mechanism for a phase transition?

It is precisely this question that motivates the present work. It acquires a particularly sharp meaning in the context of geometric microcanonical thermodynamics~\cite{di2025geometric,di2025phase,di2022geometrictheory}, where the relation between thermodynamics and geometry is not assumed in a heuristic way but can be established exactly. In that framework, the microcanonical entropy is determined by the area measure of the constant-energy shell $\Sigma_E$, and its derivatives, namely the fundamental thermodynamic observables, can be expressed as averages of local geometric quantities defined on $\Sigma_E$. The energy shell is therefore not merely the set where Hamiltonian dynamics takes place, nor the simple domain on which an equilibrium measure is induced, but the geometric locus where the thermodynamic structure of the system itself becomes accessible. The broad question pursued here about the origin of phase transition is continuous with earlier efforts by Pettini, Franzosi et al. to identify a topological origin of phase transitions~\cite{franzosi2000topology,franzosi2007topology,CASETTI2000237,franzosi1999topological,PettiniBook2007,gori2022topological,pettini2019origin,gori2018topological}, which have stimulated a broad line of research on the relation between geometry, topology, and thermodynamic criticality~\cite{kastner2011phase,casetti2009kinetic,casetti2006nonanalyticities,mehta2012energy,baroni2024simplified,angelani2003topological,angelani2005topology,baroni2005topological,kastner2008phase,risau2005topology,dicairano2021topological,di2021topology,Baroni2011,BaroniPRE2020}. The present framework, however, addresses the problem from a different angle, namely through the intrinsically microcanonical geometry of the constant-energy shell. 

In this perspective, the extrinsic curvature of the shell plays a particularly natural role, since it enters directly into microcanonical observables through the trace of the Weingarten operator, $\text{Tr} W_{\bm\xi}$. This quantity  provides a measure of the local mean curvature of $\Sigma_E$: the eigenvalues of $W_{\bm\xi}$ are the principal curvatures, namely the elementary modes according to which the shell bends along its tangent directions. If criticality reflects a loss of rigidity or a structural reorganization of the energy manifold, then it is natural to expect that its signature should appear precisely in such local geometric quantities.

\subsection*{The universal mechanism for mean-field rotor models}

The central result is to prove this intuition in a compact and universal form. For the class of mean-field rotor Hamiltonians considered here, the onset of criticality is encoded in a finite-dimensional collective geometric structure on the constant-energy shell.

\medskip
\noindent
More precisely, consider a mean-field rotor Hamiltonian of the form
\begin{equation}
H(\theta,p)
=
\sum_{i=1}^N \frac{p_i^2}{2}
+
N v_0
-
\frac{N}{2}\sum_{a,b=1}^r J_{ab}\,m_a m_b,
\end{equation}
where the collective modes \(m_a = N^{-1}\sum_{i=1}^N \phi_a(\theta_i)\) are built from an arbitrary collection of bounded trigonometric functions such as sine and cosine functions. Then, in a neighborhood of a reference symmetry branch, the Weingarten trace can be expanded in powers of the collective order parameters, and up to quadratic order its dependence is exhausted by the corresponding finite-dimensional collective sector. One thus finds
\begin{equation}
\frac{\text{Tr} W_{\bm\xi}}{N}
=
\frac{1}{c(\varepsilon)}
+
\bm m^{\mathsf T}\,\mathcal C(\varepsilon)\,\bm m
+
O(\|\bm m\|^3),
\label{eq:intro_TrW_structure}
\end{equation}
where $c(\varepsilon) = K/N$ denotes the specific kinetic energy on the reference 
branch, $\bm m$ is the vector of collective order parameters and $\mathcal C(\varepsilon)$ is a symmetric energy-dependent quadratic form, which we shall refer to as the \emph{collective curvature form}. The structural content of this expression is central: the critical threshold is not encoded in an arbitrary collection of model-dependent coefficients, but in the spectrum of a geometrically defined collective operator. The relevant critical channels are selected by the eigenvectors of $\mathcal C(\varepsilon)$, while the critical energies are determined by the vanishing of the corresponding eigenvalues. In this sense, the transition is inscribed in the geometry of the energy shell: the reorganization of the thermodynamic phase is mirrored by a reorganization of the local collective geometry of $\Sigma_E$.

The conceptual significance of this result is twofold. On the one hand, it suggests that the thermodynamic-statistical description of criticality, while remaining fully valid, may not coincide with the deepest level at which the phenomenon is organized. On the other hand, it shows that, at least for a broad and natural family of systems, the mechanism leading to the transition can be formulated in geometric terms with universal character. The point is therefore not simply to add a new criterion to the list of available tools for locating a critical threshold, but to shift the question itself: not only where the transition appears, but what structure of the system makes it possible.

\subsection*{Why the microcanonical formalism is necessary}

If the goal is to identify a structural mechanism of criticality, then the constant-energy shell must be treated as the primary equilibrium object, and the microcanonical ensemble is the natural framework in which this can be done. Unlike the canonical ensemble, the microcanonical representation provides the only thermodynamic description that remains well-defined even in regimes where canonical descriptions become ambiguous or fail---a situation made rigorous by the theory of ensemble inequivalence~\cite{touchette2005nonequivalent,touchette2004introduction,touchette2003equivalence,ellis2002nonequivalent,ellis2004thermodynamic}.

This point becomes especially important in long-range and mean-field systems that are notorious for manifesting ensemble inequivalence~\cite{barre2001inequivalence,dauxois2000violation,leyvraz2002ensemble,Pikovsky2014EnsembleInequivalence}; thus, showing that the canonical ensemble cannot always be taken as conceptually primary. At the same time, several of the assumptions that underlie standard renormalization-group approaches --- most notably scale separation and the central role of local coarse graining --- become less transparent or harder to interpret in genuinely mean-field/long-range regimes. For this reason, long-range interacting systems provide a particularly stringent testing ground for any attempt to formulate the mechanism of phase transitions at a more structural level.

It is precisely for this reason that the present work takes aim at a class of mean-field rotor Hamiltonians. These systems are not merely analytically convenient examples. They represent a regime in which the interplay between collective order, finite-size structure, and ensemble inequivalence is both physically relevant and conceptually demanding. Showing that the geometric framework developed here remains predictive in such a setting is therefore a substantive test of its scope. If a structural geometric mechanism of criticality can be identified and controlled precisely in models of this kind, then the approach is doing more than rephrasing standard thermodynamic information: it is operating in a regime where the microcanonical level is genuinely primary and where canonical intuition is not automatically sufficient.

At the same time, identifying the mechanism that underlies the onset of criticality does not by itself determine the order of the transition. Mechanism and classification are distinct aspects. The present work addresses the former by isolating the geometric structure through which criticality emerges. The latter, however, requires an additional analysis of the entropy and its derivatives. In the one-dimensional mean-field XY case, this further step was carried out explicitly in our previous works~\cite{di2025geometric,di2025phase}, where the geometric reconstruction of microcanonical observables was combined with the analysis of the corresponding critical signatures. More generally, a systematic framework for classifying phase transitions beyond the thermodynamic limit is provided by the microcanonical inflection-point analysis (MIPA) developed by Bachmann and collaborators \cite{schnabel2011microcanonical,bachmann2014novel,qi2018classification,koci2017subphase,koci2015confinement,sitarachu2020exact,sitarachu2022evidence,sitarachu2020phase}. MIPA represents a coherent and systematic method to classify phase transitions in the microcanonical ensemble based on the precise change in the energy behavior of entropy and its derivatives. In effect, MIPA represents the final step of a long study initiated by D.H.E. Gross et al.~\cite{gross2005microcanonical,gross2002geometric,gross2001ensemble,pleimling2005microcanonical,behringer2006continuous,chomaz2006challenges,chomaz1999energy,gulminelli1999critical} that identified phase transitions through convex intruders in the entropy profile. MIPA is now applied to a wide range of physical systems, such as the Potts model~\cite{Wang2024PottsFiniteSize,Liu2025PottsGeometry}, $Z(N)$-clock models~\cite{Shi2026SixStateClock}, and other spin systems~\cite{Liu2022IsingBaxterWu,Liu2025BlumeCapel}, as well as in lattice field theories~\cite{bel2021geometrical,di2022geometrictheory}, glass-transitions~\cite{vesperini2025glass}, and long-range systems~\cite{di2025geometric,di2025phase}. Interestingly, MIPA revealed to be applicable at arbitrary system sizes and remains valid also in the thermodynamic limit~\cite{di2026criticality}. In this sense, the present geometric theory should be viewed as complementary to MIPA: geometry identifies the structural mechanism, while the order of the transition is to be understood through the corresponding microcanonical thermodynamic observables.

\section{The Geometric Microcanonical Framework}
\label{sec:geometric_framework}

In this section, we briefly recall the geometric microcanonical framework developed in Refs.~\cite{di2025geometric,di2025phase} in the minimal form needed for the present work. The aim is not to rederive the whole construction, but to fix the geometric objects that enter the geometric analysis of criticality.

\subsection{Energy shells, transverse flow, and induced microcanonical measure}

Consider the Hamiltonian function defined on the phase space $\Lambda$:
\begin{equation}
H(p,q)=\sum_{i=1}^N \frac{p_i^2}{2}+V(q)
\end{equation}
and denote the constant-energy hypersurface with:
\begin{equation}
\Sigma_E=\{\bm x=(p,q)\in\Lambda:\,H(\bm x)=E\}\,.
\end{equation}
The Hamiltonian system $(\Lambda,H)$ is usually symplectic, namely, equipped with a symplectic 2-form:
\begin{equation}
    \omega:=\sum_{i=1}^N dp_i\wedge dq_i\,,
\end{equation}
which determines the Liouville volume form:
\begin{equation}
    d\mu_\Lambda:=\frac{\omega^N}{N!}=\prod_{i=1}^N dp_idq_i\,,
\end{equation}
and the Hamiltonian vector field $\bm X_H:=\partial_{p_i}H\,\partial_{q^i}-\partial_{q^i}H\,\partial_{p_i}$, which in turn provides the Hamilton equations of motion:  
\begin{equation}
   \frac{d \bm x}{dt}(t)=\bm X_H(\bm x(t)) \qquad\Leftrightarrow \qquad \left\{\begin{array}{c}
       \dfrac{dq_i}{dt}=\dfrac{\partial H}{\partial p_i}\,,\\
        \dfrac{dp_i}{dt}=-\dfrac{\partial H}{\partial q_i}\ . 
   \end{array}\right.
\end{equation}
The Hamiltonian vector field $\bm{X}_H$ is the vector that generates the Hamiltonian flow $\Phi^{\rm sym}_{t\to t'}:\Sigma_E\to\Sigma_E$, and therefore, it always lies on \(\Sigma_E\). However, the symplectic structure does not, by itself, provide a notion of normal direction, and this implies that we cannot define an intrinsic surface measure on \(\Sigma_E\). For this, one introduces a compatible Riemannian metric
\begin{equation}
    \eta=\sum_{ij}\delta_{ij}dp_i\otimes dp_j+\sum_{ij}\delta_{ij}dq_i\otimes dq_j\,,
\end{equation}
which defines the gradient \(\nabla_\eta H\) through
\begin{equation}
\eta(\nabla_\eta H,\cdot)=dH(\cdot).
\end{equation}
From a simple relation, $\eta(\nabla_\eta H,\bm X_H)=dH(\bm X_H)=\omega(\bm X_H,\bm X_H)=0$, we realize that the gradient of $H$ is always orthogonal to $\bm X_H$, namely, $\nabla_\eta H \perp \bm X_H$.

However, the appropriate transverse vector field is not the unit normal $\nabla_\eta H$, but the energy-flow generator
\begin{equation}
    \bm\xi := \frac{\nabla_\eta H}{\|\nabla_\eta H\|_\eta^2}, \qquad    dH(\bm\xi)=1.
\label{eq:xi_energy_flow}
\end{equation}
By construction, \(\bm\xi\) generates another kind of flow, $\Phi^{\rm diff}_{\bm\xi}:\Sigma_E\to\Sigma_{E'}$, that we call thermodynamic motion and maps each point of \(\Sigma_E\) to the corresponding point on \(\Sigma_{E'}\) with a uniform increment in energy. Suppose now that a local chart $\{y^\alpha\}_{\alpha=1}^{2N-1}$ is provided on $\Sigma_E$; then the Riemannian area form on $\Sigma_{E}$ , denoted by $d\sigma_E^\eta$, is defined by:
\begin{equation}
  d\sigma_E^\eta = \sqrt{\det \sigma_E^\eta}\,dy^1\cdots dy^{2N-1}\,,
\end{equation} 
and the induced microcanonical measure is then the Gelfand-Leray form
\begin{equation}
d\mu_E^\eta
=
\iota_{\bm\xi} d\mu_\Lambda
=
\frac{d\sigma_E^\eta}{\|\nabla_\eta H\|_\eta},\qquad \text{where}\quad (\iota_{\bm{\xi}}d\mu_\Lambda)(\bm{Y}_1,\ldots,\bm Y_{2N-1}) = d\mu_\Lambda(\bm{\xi},\bm{Y}_1,\ldots,\bm Y_{2N-1})\,,
\label{eq:micro_measure_geom}
\end{equation}
for all vectors $\bm{Y}_1,\ldots,\bm Y_{2N-1}$ tangent to $\Sigma_E$. Then, we see that this induced measure coincides with the Boltzmann density of states after applying the coarea formula:
\begin{equation}
    \Omega^\eta(E)=\int_{\Sigma_E} d\mu_E^\eta=\int_{\Sigma_E} \frac{d\sigma_E^\eta}{\|\nabla_\eta H\|_\eta}=\int_{\Sigma_E} \delta(H(\bm x)-E)~d\mu_{\Lambda}(\bm x)=\Omega_{\rm Boltz}(E)\,.
    \label{eqn:omega-eta_boltzmann}
\end{equation}
By the coarea formula, \(\Omega^\eta(E)\) coincides with the Boltzmann density of states.

\subsection{Unit-normal gauge and geometric entropy}

The Hamiltonian flow preserves the Liouville volume form $d\mu_\Lambda$ and, consequently, the induced measure $d\mu^\eta_E$ on each energy shell. This conservation law constrains the microcanonical weight to be of the form $d\sigma/\|\nabla H\|$ but does not constrain the metric tensor itself. This means that there exists an entire equivalence class
$[\eta] = \{g : g \sim \eta\}$ consisting of all metrics that induce the same microcanonical measure. In other words, a Riemannian metric $g$ is \textit{thermodynamically equivalent to $\eta$}, denoted $g \sim \eta$, if and only if
\begin{equation}\label{eq:gauge-equivalence-def}
    \frac{d\sigma^g_E}{\|\nabla_{\!g}H\|_g} \simeq \frac{d\sigma^\eta_E}{\|\nabla_{\!\eta} H\|_\eta}\,.
\end{equation}
In so doing, all microcanonical observables---$S(E)$, and its energy derivatives---depend only on the equivalence class $[\eta]$, not on which representative is chosen. The Euclidean metric $\eta$ plays the role of a \textit{reference gauge}: it is the canonical choice that resolves the symplectic obstruction, but any $g \in [\eta]$ describes the same thermodynamics. This freedom is a \emph{geometric gauge symmetry}. Therefore, there exists an equivalent representative of the metric class, the unit-normal gauge, in which the transverse generator has unit norm. In this gauge, one has:
\begin{equation}
\|\bm\xi\|_g=1,
\qquad
\bm\xi\equiv \nabla_g H \qquad \implies\qquad \frac{d\sigma_E^\eta}{\|\nabla_\eta H\|_\eta}= \,d\sigma_E^g\,.
\end{equation}
Therefore, the induced microcanonical measure reduces to the pure area form,
\begin{equation}
d\mu_E^g=d\sigma_E^g.
\end{equation}
Accordingly, the entropy becomes
\begin{equation}
S(E)=\ln \operatorname{area}_g(\Sigma_E)=\ln \Omega_{\rm Boltz}(E)=S_{\rm Boltz}(E)\,,
\end{equation}
so that microcanonical thermodynamics is identified with the geometry of the energy foliation.

\subsection{Weingarten operator and geometric thermodynamics}

In the unit-normal gauge, the Weingarten operator of \(\Sigma_E\) with respect to the transverse field \(\bm\xi\) is defined by
\begin{equation}
W_{\bm\xi}(\bm X):=\nabla_{\bm X} \bm\xi,
\qquad \bm X\in T_x\Sigma_E,
\end{equation}
and its trace \(\text{Tr} W_{\bm\xi}\) is the mean curvature of the energy shell. The first derivative of the geometric entropy is then
\begin{equation}
\partial_E S(E)=\langle \text{Tr} W_{\bm\xi}\rangle_E,
\label{eq:dSdE_TrW}
\end{equation}
which makes \(\text{Tr} W_{\bm\xi}\) the fundamental local geometric observable controlling the thermodynamic response.

The quantity \(\text{Tr} W_{\bm\xi}\) is written as
\begin{equation}
    \text{Tr} W_{\bm\xi}=\text{div}_g\bm\xi=\Delta_gH\,,
\end{equation}
where $\Delta_g$ is the Laplacian in the metric tensor $g$. Interestingly, the trace of $W_{\bm\xi}$ can be written directly in the Euclidean metric tensor as
\begin{equation}
\text{Tr} W_{\bm\xi}
=
\frac{\Delta_\eta H}{\|\nabla_\eta H\|_\eta^2}
-
2\frac{\nabla_\eta H^{\mathsf T}(\text{Hess}_\eta H)\nabla_\eta H}{\|\nabla_\eta H\|_\eta^4},
\label{eq:TrW_main}
\end{equation}
This is the central object of the present paper. Our goal is to show that, for a broad class of mean-field rotor Hamiltonians, the expansion of \(\text{Tr} W_{\bm\xi}\) in the collective sector takes a universal quadratic form whose spectral structure identifies the critical energies.

\section{Collective geometric expansion of the Weingarten operator}
\label{sec:general_rotor_class}

We establish that, for a broad class of mean-field rotor Hamiltonians, phase transitions are governed by a universal geometric mechanism. In this class of systems, the onset of criticality is encoded in the local geometry of the constant-energy hypersurface \(\Sigma_E\): the transition is written in the geometry of the energy shell.

More precisely, we show that the geometric observable \(\text{Tr} W_{\bm\xi}\), which controls the local extrinsic structure of \(\Sigma_E\), reorganizes near the relevant symmetry branch into a finite-dimensional collective form. Criticality is therefore not distributed over an arbitrary set of model-dependent quantities, but organized by a universal collective curvature structure acting on the space of order parameters. The critical channels are selected by its eigenmodes, and the critical energies are determined by the vanishing of the corresponding eigenvalues.

The content of the paper is thus structural. It identifies a universality class of mean-field rotor systems in which phase transitions admit a common spectral-geometric description, and in which several known models arise as particular realizations of the same mechanism.

\subsection{Mean-field rotor Hamiltonians with quadratic-order dependence in the collective parameters}

The natural domain of this result is the class of rotor Hamiltonians whose interaction is built from a finite set of bounded collective trigonometric modes and depends quadratically on them. We consider Hamiltonians of the form
\begin{equation}
H(\theta,p)
=
\sum_{i=1}^N \frac{p_i^2}{2}
+
N v_0
-
\frac{N}{2}\sum_{a,b=1}^r J_{ab}\, m_a m_b,
\label{eq:H_general_rotor}
\end{equation}
where \(v_0\) is a constant reference energy density, \(J=J^{\mathsf T}\) is a real symmetric coupling matrix, and the collective order parameters are defined by
\begin{equation}
m_a
=
\frac{1}{N}\sum_{i=1}^N \phi_a(\theta_i),
\qquad a=1,\dots,r.
\label{eq:ma_def}
\end{equation}
The functions \(\phi_a(\theta)\) are chosen from a finite family of bounded trigonometric modes, for instance
\begin{equation}
\phi_a(\theta)\in
\{\cos(k_1\theta),\sin(k_1\theta),\dots,\cos(k_s\theta),\sin(k_s\theta)\}.
\label{eq:phi_trig_family}
\end{equation}
and the coefficients $\{k_1,\ldots,k_s\}$ characterize the model and represent the allowed harmonics. Note that $\{m_a\}_{a=1}^r$ is the set of collective parameters associated with the $a$-th specific family.

This class includes, as particular cases, the isotropic and anisotropic HMF models, as well as generalized multi-harmonic rotor models. Its importance, however, is not merely that it contains several known examples. It is the natural class in which the collective expansion of \(\text{Tr} W_{\bm\xi}\) closes and the geometric mechanism of criticality becomes universal.

Introducing the vector of collective modes
\begin{equation}
\bm  m=(m_1,\dots,m_r)^{\mathsf T},
\end{equation}
in the next section, we prove that, near the relevant reference branch, the Weingarten trace admits the universal expansion
\begin{equation}
\frac{\text{Tr} W_{\bm\xi}}{N}
=
\frac{1}{c(\varepsilon)}
+
\bm  m^{\mathsf T}\,\mathcal C(\varepsilon)\,\bm  m
+
O(\|\bm  m\|^3),
\label{eq:TrW_collective_scheme}
\end{equation}
where \(\mathcal C(\varepsilon)\) is a symmetric energy-dependent quadratic form on the collective sector. This \(\mathcal C(\varepsilon)\) is the collective curvature form that organizes the transition.

In order to show this, we use the explicit representation
\begin{equation}
\text{Tr} W_{\bm\xi}
=
\frac{\Delta_\eta H}{\|\nabla_\eta H\|_\eta^2}
-
2\,
\frac{\nabla_\eta H^{\mathsf T}(\text{Hess}_\eta H)\nabla_\eta H}{\|\nabla_\eta H\|_\eta^4}.
\label{eq:TrW_main_sec4}
\end{equation}
For the rotor Hamiltonians considered here, the leading collective structure is already fully visible in the first ratio, while the Hessian contraction contributes only at subleading order in \(N\), as shown in Appendix~\ref{app:subleading-hessian-force}. The result is a universal quadratic expansion of \(\text{Tr} W_{\bm\xi}/N\) in the collective modes.

\subsection{Differential structure of the collective sector}

The derivation of the collective expansion is somewhat lengthy and technically articulated. In the main text, we isolate the structural steps leading to the universal curvature form, while the full calculation is given in Appendix~\ref{app:collective_expansion}.

\medskip
Consider the \(\theta\)-dependent vector collecting the family modes,
\begin{equation}
\Phi(\theta)
=
\bigl(\phi_1(\theta),\dots,\phi_r(\theta)\bigr)^{\mathsf T},
\qquad
\bm m
=
\frac{1}{N}\sum_{i=1}^N \Phi(\theta_i)
=
\frac{1}{N}\sum_{i=1}^N
\begin{pmatrix}
\phi_1(\theta_i)\\
\phi_2(\theta_i)\\
\vdots\\
\phi_r(\theta_i)
\end{pmatrix}.
\label{eq:Phi_m_sec4}
\end{equation}
We consider the Hamiltonian
\begin{equation}
H(\theta,p)
=
\sum_{i=1}^N \frac{p_i^2}{2}
+
N v_0
-
\frac{N}{2}\,\bm m^{\mathsf T}J\,\bm m,
\qquad
V(\theta)
:=
N v_0
-
\frac{N}{2}\,\bm m^{\mathsf T}J\,\bm m.
\label{eq:H_sec4}
\end{equation}
For notational simplicity, we suppress the metric label \(\eta\) in what follows.

The gradient of the Hamiltonian is
\begin{equation}
\nabla H
=
\left(
\frac{\partial H}{\partial p_i},
\frac{\partial V}{\partial \theta_i}
\right)^{\mathsf T}
=
\left(
p_i,
-(J\bm m)^{\mathsf T}\Phi'(\theta_i)
\right)^{\mathsf T},
\label{eq:first_derivatives_sec4}
\end{equation}
where
\begin{equation}
\Phi'(\theta)
=
\bigl(\phi_1'(\theta),\dots,\phi_r'(\theta)\bigr)^{\mathsf T}.
\end{equation}

The Hessian of the potential is
\begin{equation}
\frac{\partial^2 V}{\partial \theta_i\,\partial \theta_j}
=
-
\frac{1}{N}\,
\Phi'(\theta_i)^{\mathsf T}J\,\Phi'(\theta_j)
-
\delta_{ij}\,(J\bm m)^{\mathsf T}\Phi''(\theta_i),
\qquad
i,j=1,\dots,N,
\label{eq:angular_Hessian_general}
\end{equation}
and therefore the Hessian of the full Hamiltonian reads
\begin{equation}
\mathrm{Hess}\,H
=
\begin{pmatrix}
\mathbb I_N & 0\\
0 & \mathrm{Hess}_{\theta}V
\end{pmatrix}.
\label{eq:Hessian_block_form}
\end{equation}
Its Laplacian is
\begin{equation}
\Delta H
:=
\mathrm{Tr}(\mathrm{Hess}\,H)
=
N+\mathrm{Tr}(\mathrm{Hess}_{\theta}V)
=
N-\sum_{i=1}^N\left[
\frac{1}{N}\,
\Phi'(\theta_i)^{\mathsf T}J\,\Phi'(\theta_i)
+
(J\bm m)^{\mathsf T}\Phi''(\theta_i)
\right].
\label{eq:DeltaH_raw_sec4}
\end{equation}

The closure of the expansion relies on a basic property of the trigonometric family: the second derivatives remain inside the same finite-dimensional mode space. More precisely, there exists a constant matrix \(D\) such that
\begin{equation}
\Phi''(\theta)=-D\,\Phi(\theta).
\label{eq:D_closure}
\end{equation}
For the standard harmonic basis
\(
\{\cos(k_\alpha\theta),\sin(k_\alpha\theta)\}_{\alpha=1}^s
\),
the matrix \(D\) is block diagonal and positive:
\begin{equation}
D
=
\mathrm{diag}\!\bigl(
k_1^2,k_1^2,\dots,k_s^2,k_s^2
\bigr).
\label{eq:D_harmonic}
\end{equation}

\subsection{Expansion of \texorpdfstring{\(\Delta H\)}{Delta H}}

Using the closure relation \eqref{eq:D_closure}, one obtains
\begin{equation}
\Delta H
=
N
-
\frac{1}{N}\sum_{i=1}^N \Phi'(\theta_i)^{\mathsf T}J\,\Phi'(\theta_i)
+
N\,(J\bm m)^{\mathsf T}D\,\bm m.
\label{eq:DeltaH_exact_sec4}
\end{equation}
Introducing the empirical matrix
\begin{equation}
Q_N(\bm m)
:=
\frac{1}{N}\sum_{i=1}^N
\Phi'(\theta_i)\Phi'(\theta_i)^{\mathsf T},
\label{eq:Qm_def}
\end{equation}
this can be rewritten as (see Appendix~\ref{app:expansion-deltaH})
\begin{equation}
\frac{\Delta H}{N}
=
1
-
\frac{1}{N}\mathrm{Tr}\!\bigl(JQ_N(\bm m)\bigr)
+
\bm m^{\mathsf T}J D\,\bm m.
\label{eq:DeltaH_over_N_preexp}
\end{equation}

We now expand around a reference branch, typically the disordered branch \(\bm m=\bm 0\), and assume that the large-\(N\) limit of \(Q_N(\bm m)\) exists and defines a smooth matrix-valued function \(Q(\bm m)\). Denoting by
\begin{equation}
Q_\ast:=Q(\bm 0)
\label{eq:Qstar_def}
\end{equation}
its value on the reference branch, one has
\begin{equation}
Q(\bm m)=Q_\ast+O(\|\bm m\|).
\label{eq:Q_branch_exp}
\end{equation}
Since \(N^{-1}\mathrm{Tr}(JQ_\ast)=O(N^{-1})\), the leading collective contribution is (see Appendix~\ref{app:expansion-deltaH-branch})
\begin{equation}
\frac{\Delta H}{N}
=
1
+
\bm m^{\mathsf T}\mathcal M\,\bm m
+
O(\|\bm m\|^3)
+
O(N^{-1}),
\label{eq:DeltaH_expanded}
\end{equation}
where the symmetric matrix \(\mathcal M\) is the symmetrized part of \(JD\),
\begin{equation}
\mathcal M
:=
\frac{1}{2}\bigl(JD+D^{\mathsf T}J\bigr).
\label{eq:M_matrix}
\end{equation}
If \(D\) is symmetric, this becomes
\begin{equation}
\mathcal M
=
\frac{1}{2}\bigl(JD+DJ\bigr)\,.
\label{eq:M_matrix_symD}
\end{equation}

Equation~\eqref{eq:DeltaH_expanded} shows that the Laplacian of the Hamiltonian acquires a collective quadratic correction entirely determined by the coupling matrix \(J\) and by the closure matrix \(D\).

\subsection{Expansion of \texorpdfstring{\(\|\nabla H\|^2\)}{||grad H||^2}}

From Eq.~\eqref{eq:first_derivatives_sec4}, the squared norm of the gradient is
\begin{equation}
\|\nabla H\|^2
=
\sum_{i=1}^N p_i^2
+
\sum_{i=1}^N
\Bigl[(J\bm m)^{\mathsf T}\Phi'(\theta_i)\Bigr]^2.
\label{eq:gradnorm_exact_sec4}
\end{equation}
Using again the matrix \(Q(\bm m)\), this becomes
\begin{equation}
\|\nabla H\|^2
=
K
+
N\,\bm m^{\mathsf T}J\,Q(\bm m)\,J\,\bm m,
\label{eq:gradnorm_exact_Q}
\end{equation}
where
\begin{equation}
K:=\sum_{i=1}^N p_i^2.
\label{eq:K_def}
\end{equation}
Dividing by \(N\), we define (see Appendix~\ref{app:expansion-G})
\begin{equation}
G:=\frac{\|\nabla H\|^2}{N}
=
c
+
\bm m^{\mathsf T}\mathcal B\,\bm m
+
O(\|\bm m\|^3),
\label{eq:G_expanded}
\end{equation}
where
\begin{equation}
c:=\frac{K}{N},
\label{eq:c_def}
\end{equation}
is the specific kinetic background on the reference branch, and
\begin{equation}
\mathcal B
:=
J\,Q_\ast\,J.
\label{eq:B_matrix}
\end{equation}

The matrix \(\mathcal B\) contains the directional stiffness induced by the collective field through the derivative covariance of the underlying trigonometric basis. It is the second universal ingredient entering the collective geometry of the shell.

For the uniform disordered branch of a harmonic basis, the matrix \(Q_\ast\) is diagonal:
\begin{equation}
Q_\ast
=
\frac{1}{2}\,
\mathrm{diag}\!\bigl(
k_1^2,k_1^2,\dots,k_s^2,k_s^2
\bigr),
\label{eq:Q_uniform_branch}
\end{equation}
so that \(\mathcal B\) is explicitly determined by the coupling matrix \(J\) and the harmonic content of the basis.

\subsubsection{Leading-order form of \texorpdfstring{\(\text{Tr} W_{\bm\xi}/N\)}{Tr Wxi/N}}

We now combine the expansions above with Eq.~\eqref{eq:TrW_main_sec4}. At leading order in the thermodynamic limit, the Hessian contraction term contributes only subleading corrections to \(\text{Tr} W_{\bm\xi}/N\) (see the derivation in Appendix~\ref{app:subleading-hessian-force}), so that:
\begin{equation}
\frac{\text{Tr} W_{\bm\xi}}{N}
=
\frac{\Delta H/N}{G}
+
O(N^{-1}).
\label{eq:TrW_reduction_sec4}
\end{equation}
Using Eqs.~\eqref{eq:DeltaH_expanded} and \eqref{eq:G_expanded}, one finds (see Appendix~\ref{app:expansion-TrW}):
\begin{equation}
\frac{\text{Tr} W_{\bm\xi}}{N}
=
\frac{1}{c}
+
\bm m^{\mathsf T}\mathcal C(\varepsilon)\,\bm m
+
O(\|\bm m\|^3)
+
O(N^{-1}),
\label{eq:TrW_final_expansion}
\end{equation}
with
\begin{equation}
\mathcal C(\varepsilon)
=
\frac{1}{c^2}
\Bigl(
c\,\mathcal M-\mathcal B
\Bigr).
\label{eq:C_general}
\end{equation}
Equivalently, one finds:
\begin{equation}
\mathcal C(\varepsilon)
=
\frac{1}{c^2}
\left[
\frac{c}{2}\bigl(JD+D^{\mathsf T}J\bigr)
-
JQ_\ast J
\right].
\label{eq:C_general_explicit}
\end{equation}

This is the universal collective curvature form depending on the energy only through the branch kinetic background \(c=c(\varepsilon)\), while its geometric structure is fixed by the coupling matrix \(J\), the closure matrix \(D\) of the trigonometric family, and the branch covariance matrix \(Q_\ast\). The critical channels are therefore selected by the eigenvectors of \(\mathcal C(\varepsilon)\), and the critical energies are identified by the vanishing of its eigenvalues.

\subsection*{Diagonal harmonic sectors}

A particularly transparent case is obtained when the collective modes are organized in harmonic pairs and the coupling matrix \(J\) is diagonal in the same basis. Denoting by \(J_\alpha\) the coupling of the \(\alpha\)-th harmonic sector and by \(k_\alpha\) the associated wave number, one has:
\begin{equation}
D_\alpha=k_\alpha^2,
\qquad
(Q_\ast)_\alpha=\frac{k_\alpha^2}{2},
\label{eq:diag_sector_data}
\end{equation}
and therefore the corresponding curvature coefficient reads:
\begin{equation}
\mathcal C_\alpha(\varepsilon)
=
\frac{k_\alpha^2 J_\alpha}{c^2}
\left(
c-\frac{J_\alpha}{2}
\right).
\label{eq:C_alpha_diag}
\end{equation}
This is the direct generalization of the coefficient found in the single-mode mean-field XY case. It shows explicitly that the critical energy of a given collective channel is determined by the vanishing of a geometric curvature coefficient.

The same logic persists in the fully coupled case, where \(J\) is not diagonal and \(\mathcal C(\varepsilon)\) must be analyzed spectrally as a matrix acting on the collective order-parameter space. This spectral interpretation is the subject of the next section.

\section{Spectral criterion for criticality}
\label{sec:spectral_criterion}

The expansion derived in the previous section shows that, for the class of mean-field rotor Hamiltonians considered here, the local geometry of the energy shell reorganizes near the reference branch into a finite-dimensional quadratic form acting on the collective sector. This result has an immediate and far-reaching consequence: criticality becomes a spectral property of the collective curvature form. Indeed, the expansion
\begin{equation}
\frac{\text{Tr} W_{\bm\xi}}{N}
=
\frac{1}{c(\varepsilon)}
+
\bm m^{\mathsf T}\,\mathcal C(\varepsilon)\,\bm m
+
O(\|\bm m\|^3)
\label{eq:TrW_sec5_start}
\end{equation}
shows that all the nontrivial geometric information relevant to the instability is encoded, at quadratic order, in the symmetric operator \(\mathcal C(\varepsilon)\). The role of the collective modes is therefore sharply separated from the background contribution: the first term fixes the regular geometric background of the shell, while the second term identifies the channels along which the geometry reorganizes at the transition.

\subsection{Eigenmodes of the collective curvature form}

Since \(\mathcal C(\varepsilon)\) is symmetric, it admits an orthonormal spectral decomposition. Let
\begin{equation}
\mathcal C(\varepsilon)\,\bm v_\alpha(\varepsilon)
=
\lambda_\alpha(\varepsilon)\,\bm v_\alpha(\varepsilon),
\qquad
\alpha=1,\dots,r,
\label{eq:C_eigenproblem}
\end{equation}
where the eigenvectors \(\bm v_\alpha\) define the distinguished collective directions in order-parameter space and the eigenvalues \(\lambda_\alpha(\varepsilon)\) measure the corresponding geometric curvature coefficients. Expanding
\begin{equation}
\bm m=\sum_{\alpha=1}^r u_\alpha\,\bm v_\alpha,
\label{eq:m_eigenbasis}
\end{equation}
Eq.~\eqref{eq:TrW_sec5_start} becomes
\begin{equation}
\frac{\text{Tr} W_{\bm\xi}}{N}
=
\frac{1}{c(\varepsilon)}
+
\sum_{\alpha=1}^r \lambda_\alpha(\varepsilon)\,u_\alpha^2
+
O(\|\bm m\|^3).
\label{eq:TrW_eigenbasis}
\end{equation}

This representation is the geometric normal form of the transition. It shows that the energy shell does not reorganize arbitrarily near criticality: it does so along a finite set of collective eigendirections selected by the spectrum of \(\mathcal C(\varepsilon)\).

\subsection{Vanishing eigenvalues as geometric critical conditions}

The quadratic expansion \eqref{eq:TrW_eigenbasis} identifies the leading geometric response of the energy shell in each collective direction. A critical channel is reached when one of the curvature coefficients \(\lambda_\alpha(\varepsilon)\) vanishes. At that point, the quadratic geometric stiffness along the corresponding eigendirection disappears, and the structure of the energy shell becomes marginal in that collective channel.

This leads to the geometric critical condition
\begin{equation}
\lambda_\alpha(\varepsilon_c)=0
\label{eq:critical_condition_lambda}
\end{equation}
for at least one \(\alpha\). Equivalently, criticality is characterized by
\begin{equation}
\det \mathcal C(\varepsilon_c)=0,
\label{eq:critical_condition_det}
\end{equation}
or, more precisely, by the vanishing of the eigenvalue associated with the physically relevant instability branch.

Equation~\eqref{eq:critical_condition_lambda} is the local geometric signature of the transition. It expresses the onset of criticality as the disappearance of a restoring curvature in the collective geometry of \(\Sigma_E\). In this sense, the transition is encoded in the energy shell itself: it occurs when the collective extrinsic geometry becomes marginal along a distinguished mode.

\subsection{Critical energies from the spectrum of \texorpdfstring{\(J\)}{J}}

The previous section showed that
\begin{equation}
\mathcal C(\varepsilon)
=
\frac{1}{c(\varepsilon)^2}
\bigl(
c(\varepsilon)\,\mathcal M-\mathcal B
\bigr),
\label{eq:C_again}
\end{equation}
with \(\mathcal M\) and \(\mathcal B\) determined by the harmonic structure and the coupling matrix \(J\). In the particularly transparent case in which the collective basis diagonalizes the relevant harmonic sectors, one has
\begin{equation}
\mathcal C_\alpha(\varepsilon)
=
\frac{k_\alpha^2 J_\alpha}{c(\varepsilon)^2}
\left(
c(\varepsilon)-\frac{J_\alpha}{2}
\right),
\label{eq:C_alpha_repeat}
\end{equation}
so that the critical condition becomes
\begin{equation}
c(\varepsilon_{c,\alpha})=\frac{J_\alpha}{2}.
\label{eq:c_condition}
\end{equation}

For the standard rotor branch, where
\begin{equation}
c(\varepsilon)=2(\varepsilon-v_0),
\label{eq:c_eps_relation}
\end{equation}
this yields
\begin{equation}
\varepsilon_{c,\alpha}
=
v_0+\frac{J_\alpha}{4}.
\label{eq:epsc_alpha}
\end{equation}

Equation~\eqref{eq:epsc_alpha} is the simplest explicit manifestation of the spectral-geometric criterion: the critical energies are fixed by the collective spectrum of the interaction matrix, with the harmonic content entering through the geometric coefficients of the curvature expansion.

In the fully coupled case, where \(J\) is not diagonal in the chosen collective basis (see Section~\ref{sec:non-diagonal-sector}), the same logic survives in spectral form. The critical energies are then determined by the vanishing of the eigenvalues of \(\mathcal C(\varepsilon)\), and hence by the interplay between the collective coupling structure encoded in \(J\) and the geometric tensors \(D\) and \(Q_\ast\).

\subsection{Universal geometric criterion for criticality}

We may summarize the result of the previous sections in the following statement.

\medskip

\noindent
\noindent
\textbf{Proposition.}
\emph{Consider a mean-field rotor Hamiltonian of the form}
\begin{equation}
H(\theta,p)
=
\sum_{i=1}^N \frac{p_i^2}{2}
+
N v_0
-
\frac{N}{2}\sum_{a,b=1}^r J_{ab}\,m_a m_b,
\end{equation}
\emph{where the collective modes} \(m_a = N^{-1}\sum_{i=1}^N \phi_a(\theta_i)\) \emph{are built from a finite family of bounded trigonometric functions. Then, near a reference symmetry branch,} \(\text{Tr} W_{\bm\xi}\) \emph{admits an expansion in the collective order parameters} \(\bm m\), \emph{and up to quadratic order, this dependence is entirely controlled by the associated finite-dimensional collective sector. In particular,}
\begin{equation}
\frac{\text{Tr} W_{\bm\xi}}{N}
=
\frac{1}{c(\varepsilon)}
+
\bm m^{\mathsf T}\,\mathcal C(\varepsilon)\,\bm m
+
O(\|\bm m\|^3),
\end{equation}
\emph{where \(\mathcal C(\varepsilon)\) is a symmetric energy-dependent collective curvature form. The critical channels are the eigenmodes of \(\mathcal C(\varepsilon)\), and the critical energies are determined by the vanishing of its eigenvalues.
}

\medskip

This proposition is the central structural result of the paper. It shows that, within this natural class of mean-field rotor systems, criticality is governed by a universal geometric mechanism. Different models correspond to different realizations of the same collective curvature structure.

\subsection*{Physical interpretation}

The spectral criterion above gives a precise meaning to the statement that phase transitions are written in the geometry of the energy shell. The transition is not identified here through a singularity of a thermodynamic potential, nor through a self-consistency equation taken in isolation, but through the reorganization of the local extrinsic geometry of \(\Sigma_E\) in the collective sector.

The background contribution \(1/c(\varepsilon)\) describes the regular geometric scale of the shell. By contrast, the operator \(\mathcal C(\varepsilon)\) isolates the directions in order-parameter space along which this geometry becomes unstable or marginal. The transition occurs when one of these collective curvature coefficients vanishes, signaling that the shell loses quadratic geometric rigidity along a specific collective mode.

From this viewpoint, known mean-field critical energies are not isolated algebraic facts tied to the details of individual models. They are spectral-geometric corollaries of a common mechanism. This is precisely the sense in which the present framework identifies a universality class of geometric critical behavior.

\section{Recovery of known and generalized mean-field rotor models}
\label{sec:recovery_models}

We now show that the spectral-geometric criterion derived above reproduces the critical structure of several standard mean-field rotor models. This is an essential consistency check, but also something more: it shows that the models usually treated separately are in fact particular realizations of the same collective curvature mechanism. Their critical energies are not isolated results obtained case by case, but corollaries of a common geometric structure.

\subsection{Isotropic HMF model}

The isotropic Hamiltonian mean-field model is defined by~\cite{antoni1995clustering}
\begin{equation}
H
=
\sum_{i=1}^N \frac{p_i^2}{2}
+
\frac{1}{2N}\sum_{i,j=1}^N \bigl[1-\cos(\theta_i-\theta_j)\bigr].
\label{eq:H_HMF_standard}
\end{equation}
Introducing the usual magnetization components
\begin{equation}
m_x=\frac{1}{N}\sum_{i=1}^N \cos\theta_i,
\qquad
m_y=\frac{1}{N}\sum_{i=1}^N \sin\theta_i,
\label{eq:mx_my_HMF}
\end{equation}
the potential can be written as
\begin{equation}
V
=
\frac{N}{2}
-
\frac{N}{2}\left(m_x^2+m_y^2\right).
\label{eq:V_HMF_mxy}
\end{equation}
Hence the model belongs to the class \eqref{eq:H_general_rotor} with
\begin{equation}
v_0=\frac{1}{2},
\qquad
\bm m=(m_x,m_y)^{\mathsf T},
\qquad
J=\mathbb I_2.
\label{eq:HMF_data}
\end{equation}

The relevant trigonometric basis is the \(k=1\) harmonic sector,
\begin{equation}
\Phi(\theta)=
\begin{pmatrix}
\cos\theta\\
\sin\theta
\end{pmatrix},
\qquad
D=\mathbb I_2,
\qquad
Q_\ast=\frac{1}{2}\mathbb I_2
\label{eq:HMF_basis_data}
\end{equation}
on the disordered branch. Therefore
\begin{equation}
\mathcal C(\varepsilon)
=
\frac{1}{c(\varepsilon)^2}
\left(c(\varepsilon)-\frac{1}{2}\right)\mathbb I_2,
\label{eq:C_HMF}
\end{equation}
and the critical condition reads
\begin{equation}
c(\varepsilon_c)=\frac{1}{2}.
\label{eq:c_HMF}
\end{equation}
Using \(c(\varepsilon)=2(\varepsilon-v_0)=2\varepsilon-1\), one obtains
\begin{equation}
\varepsilon_c=\frac{3}{4},
\label{eq:epsc_HMF}
\end{equation}
which is the standard critical energy of the isotropic HMF model, which coincides with the results obtained in Refs.~\cite{di2025geometric,di2025phase}.

In this case, the collective curvature form is proportional to the identity. The two collective directions \(m_x\) and \(m_y\) are therefore equivalent, reflecting the rotational symmetry of the model. Criticality corresponds to the simultaneous geometric softening of the two symmetry-related directions.

\subsection{Anisotropic HMF model}

The first extension of the HMF model is the anisotropic XY model defined with the Hamiltonian~\cite{Mukamel2008NotesLongRange}:
\begin{equation}
H
=
\sum_{i=1}^N \frac{p_i^2}{2}
+
\frac{1}{2N}\sum_{i,j=1}^N \bigl[1-\cos(\theta_i-\theta_j)\bigr]
-
\frac{D_a}{2N}\left(\sum_{i=1}^N \cos\theta_i\right)^2,
\label{eq:H_aniso}
\end{equation}
where \(D_a\) denotes the anisotropy parameter. In terms of \((m_x,m_y)\), the potential becomes
\begin{equation}
V
=
\frac{N}{2}
-
\frac{N}{2}\Bigl[(1+D_a)m_x^2+m_y^2\Bigr].
\label{eq:V_aniso}
\end{equation}
Thus the model is again of the form \eqref{eq:H_general_rotor}, with
\begin{equation}
v_0=\frac{1}{2},
\qquad
J=
\begin{pmatrix}
1+D_a & 0\\
0 & 1
\end{pmatrix}.
\label{eq:J_aniso}
\end{equation}

Since the basis and the branch covariance are unchanged,
\begin{equation}
D=-\mathbb I_2,
\qquad
Q_\ast=\frac{1}{2}\mathbb I_2,
\end{equation}
the collective curvature form is diagonal,
\begin{equation}
\mathcal C(\varepsilon)
=
\frac{1}{c(\varepsilon)^2}
\begin{pmatrix}
(1+D_a)\left(c(\varepsilon)-\frac{1+D_a}{2}\right) & 0\\
0 & c(\varepsilon)-\frac{1}{2}
\end{pmatrix}.
\label{eq:C_aniso}
\end{equation}
The two collective channels are no longer equivalent. Their vanishing conditions are
\begin{equation}
c(\varepsilon_{c,x})=\frac{1+D_a}{2},
\qquad
c(\varepsilon_{c,y})=\frac{1}{2},
\end{equation}
which give
\begin{equation}
\varepsilon_{c,x}=\frac{3+D_a}{4},
\qquad
\varepsilon_{c,y}=\frac{3}{4}.
\label{eq:epsc_aniso}
\end{equation}

For \(D_a>0\), the physically relevant instability develops in the \(x\)-channel, and the equilibrium transition is therefore controlled by
\begin{equation}
\varepsilon_c=\frac{3+D_a}{4},
\label{eq:epsc_aniso_phys}
\end{equation}
in agreement with the known result. The isotropic value \(3/4\) remains as the geometric threshold of the orthogonal channel, which is no longer the dominant instability.

This example makes the meaning of the formalism especially transparent: anisotropy does not change the geometric mechanism itself, but splits the collective curvature spectrum and thereby selects a preferred critical direction.

\subsection{Generalized HMF model}

We next consider the generalized HMF model with first and second harmonics~\cite{Pikovsky2014EnsembleInequivalence},
\begin{equation}
H
=
\sum_{i=1}^N \frac{p_i^2}{2}
+
\frac{N}{2}
-
\frac{N}{2}\Bigl[\Delta\,m_1^2+(1-\Delta)\,m_2^2\Bigr],
\label{eq:H_GHMF}
\end{equation}
where
\begin{equation}
m_1=\frac{1}{N}\sum_{i=1}^N \cos\theta_i,
\qquad
m_2=\frac{1}{N}\sum_{i=1}^N \cos 2\theta_i.
\label{eq:R1_R2_GHMF}
\end{equation}
At the level of the disordered branch and of the leading quadratic collective expansion, the relevant sector is diagonal in the first and second harmonic amplitudes. The model is therefore described by
\begin{equation}
v_0=\frac{1}{2},
\qquad
J=
\begin{pmatrix}
\Delta & 0\\
0 & 1-\Delta
\end{pmatrix},
\label{eq:J_GHMF}
\end{equation}
with harmonic numbers
\begin{equation}
k_1=1,
\qquad
k_2=2.
\label{eq:k_GHMF}
\end{equation}

Using Eq.~\eqref{eq:C_alpha_repeat}, one obtains
\begin{equation}
\mathcal C_1(\varepsilon)
=
\frac{\Delta}{c(\varepsilon)^2}
\left(c(\varepsilon)-\frac{\Delta}{2}\right),
\label{eq:C1_GHMF}
\end{equation}
and
\begin{equation}
\mathcal C_2(\varepsilon)
=
\frac{4(1-\Delta)}{c(\varepsilon)^2}
\left(c(\varepsilon)-\frac{1-\Delta}{2}\right).
\label{eq:C2_GHMF}
\end{equation}
The corresponding vanishing conditions yield
\begin{equation}
\varepsilon_c^{(1)}
=
\frac{1}{2}+\frac{\Delta}{4},
\qquad
\varepsilon_c^{(2)}
=
\frac{1}{2}+\frac{1-\Delta}{4}
=
\frac{3-\Delta}{4}.
\label{eq:epsc_GHMF}
\end{equation}

These are precisely the critical energies associated with the ferromagnetic and nematic channels on the disordered branch. 

At the level of the disordered branch, the GHMF model is recovered in the precise sense that the geometric splitting of the first and second harmonic channels reproduces the \(P\to F\) and \(P\to N\) critical lines. This, however, does not exhaust the full three-phase structure of the model. In particular, the \(N\to F\) transition does not originate from the disordered configuration \((R_1,R_2)=(0,0)\), but from a symmetry-broken nematic background with \(R_1=0\) and \(R_2=R_2^*\neq 0\). Capturing this transition therefore requires a second expansion of \(\text{Tr} W_{\bm\xi}\), now performed with respect to fluctuations of the ferromagnetic order parameter \(R_1\) around the nematic branch. In that case, the resulting geometric criterion acquires an implicit form through the branch-dependent background \(R_2^*(\varepsilon)\). Since the present work is focused on the universal structure of the criterion and on its explicit realization at the level of the disordered branch, this second expansion will be developed separately in a dedicated work of the GHMF model~\cite{GHMFinprep}.

The importance of this example is conceptual as well as technical. The two critical energies are not obtained here from a model-specific self-consistency calculation, but from the splitting of the collective geometric spectrum into distinct harmonic channels. The first and second harmonics correspond to different eigendirections of the curvature form, and each critical line is the energy at which the corresponding geometric stiffness vanishes.

\subsection{Multi-harmonic generalized HMF model}
\label{subsec:multi_harmonic_GHMF}

We now analyze a natural extension of the GHMF model. This model defines a natural multi-harmonic extension of the HMF/GHMF class, in which a finite set of collective Fourier modes competes at the mean-field level:
\begin{equation}
H
=
\sum_{i=1}^N \frac{p_i^2}{2}
+
\frac{1}{2N}\sum_{i,j=1}^N
\left[
1-\sum_{k=1}^{K}\Delta_k \cos\bigl(k(\theta_i-\theta_j)\bigr)
\right],
\label{eq:H_multiGHMF}
\end{equation}
with
\begin{equation}
\sum_{k=1}^{K}\Delta_k=1,
\qquad
\Delta_k\ge 0.
\label{eq:Delta_sum_multi}
\end{equation}
Using the trigonometric identity
\begin{equation}
\cos\bigl(k(\theta_i-\theta_j)\bigr)
=
\cos(k\theta_i)\cos(k\theta_j)
+
\sin(k\theta_i)\sin(k\theta_j),
\label{eq:cos_diff_identity}
\end{equation}
the potential can be written as
\begin{equation}
V
=
\frac{N}{2}
-
\frac{N}{2}\sum_{k=1}^{K}\Delta_k
\left(
m_{c,k}^2+m_{s,k}^2
\right),
\label{eq:V_multiGHMF}
\end{equation}
where
\begin{equation}
m_{c,k}
=
\frac{1}{N}\sum_{i=1}^N \cos(k\theta_i),
\qquad
m_{s,k}
=
\frac{1}{N}\sum_{i=1}^N \sin(k\theta_i).
\label{eq:mc_ms_multi}
\end{equation}

Equivalently, introducing the usual harmonic order parameters
\begin{equation}
R_k
=
\frac{1}{N}\left|\sum_{i=1}^N e^{ik\theta_i}\right|,
\label{eq:Rk_def}
\end{equation}
one has
\begin{equation}
R_k^2=m_{c,k}^2+m_{s,k}^2,
\label{eq:Rk_mc_ms}
\end{equation}
and therefore
\begin{equation}
V
=
\frac{N}{2}
-
\frac{N}{2}\sum_{k=1}^{K}\Delta_k R_k^2.
\label{eq:V_multiGHMF_Rk}
\end{equation}

This model belongs to the general class \eqref{eq:H_general_rotor} with
\begin{equation}
v_0=\frac{1}{2},
\label{eq:v0_multi}
\end{equation}
and with a collective basis formed by the harmonic pairs
\begin{equation}
\Phi_k(\theta)
=
\begin{pmatrix}
\cos(k\theta)\\
\sin(k\theta)
\end{pmatrix},
\qquad
k=1,\dots,K.
\label{eq:Phi_k_multi}
\end{equation}
On the disordered branch, each harmonic sector is diagonal and characterized by
\begin{equation}
D_k=-k^2\mathbb I_2,
\qquad
(Q_\ast)_k=\frac{k^2}{2}\mathbb I_2.
\label{eq:DQ_multi}
\end{equation}
The collective curvature coefficients are therefore
\begin{equation}
\mathcal C_k(\varepsilon)
=
\frac{k^2\Delta_k}{c(\varepsilon)^2}
\left(
c(\varepsilon)-\frac{\Delta_k}{2}
\right),
\label{eq:Ck_multi}
\end{equation}
with the same coefficient for the cosine and sine components of the \(k\)-th harmonic.

The corresponding geometric critical condition is
\begin{equation}
c(\varepsilon_{c,k})=\frac{\Delta_k}{2},
\label{eq:c_multi_crit}
\end{equation}
which, using \(c(\varepsilon)=2(\varepsilon-v_0)=2\varepsilon-1\), gives
\begin{equation}
\varepsilon_{c,k}
=
\frac{1}{2}+\frac{\Delta_k}{4},
\qquad
k=1,\dots,K.
\label{eq:epsc_multi}
\end{equation}

The standard GHMF model is recovered as the case \(K=2\), with
\begin{equation}
\Delta_1=\Delta,
\qquad
\Delta_2=1-\Delta.
\label{eq:GHMF_as_multi}
\end{equation}
Equation~\eqref{eq:epsc_multi} then reduces to
\begin{equation}
\varepsilon_c^{(1)}=\frac{1}{2}+\frac{\Delta}{4},
\qquad
\varepsilon_c^{(2)}=\frac{1}{2}+\frac{1-\Delta}{4}
=
\frac{3-\Delta}{4},
\label{eq:epsc_multi_to_GHMF}
\end{equation}
in agreement with Eq.~\eqref{eq:epsc_GHMF}.

This example is important because it makes the universality class completely explicit. The GHMF model with first and second harmonics is not an isolated case, but the first nontrivial member of a broader multi-harmonic family. In all these models, the transition is governed by the same mechanism: each harmonic sector carries its own collective curvature coefficient, and the corresponding critical line is the energy at which that geometric stiffness vanishes.

\subsection{Non-diagonal collective sectors}
\label{sec:non-diagonal-sector}
The previous examples all correspond to situations in which the collective basis diagonalizes the relevant sector of the interaction. In that case the critical channels coincide directly with the chosen collective variables, and the interpretation of the transition is particularly transparent.

The general formalism, however, is not restricted to diagonal sectors. Whenever the coupling matrix \(J\) is not diagonal in the chosen collective basis, the collective curvature form \(\mathcal C(\varepsilon)\) acquires off-diagonal entries, and the physically relevant critical modes are no longer the original order parameters themselves, but the eigenvectors of \(\mathcal C(\varepsilon)\). The transition is then organized by mixed collective directions.

This distinction is important. The geometric mechanism is the same in both cases, but diagonal models hide part of its content by making the eigendirections obvious from the start. The non-diagonal case reveals more clearly that the fundamental object is not the chosen order-parameter basis, but the spectral structure of the collective curvature form.

The next section develops this point further by considering non-diagonal collective sectors, where the critical channels are genuinely mixed and must be identified through the full spectral resolution of the collective curvature operator.

\subsection{A non-diagonal anisotropic HMF model}
\label{subsec:nondiag_HMF}

A natural non-diagonal extension of the anisotropic HMF model is obtained by allowing an explicit mixing between the \(\cos\theta\) and \(\sin\theta\) sectors already at the level of the two-body interaction. We consider the Hamiltonian
\begin{equation}
H
=
\sum_{i=1}^N \frac{p_i^2}{2}
+
\frac{N}{2}
-
\frac{1}{2N}\sum_{i,j=1}^N
\Big[
J_x \cos\theta_i \cos\theta_j
+
J_{xy}\bigl(\cos\theta_i\sin\theta_j+\sin\theta_i\cos\theta_j\bigr)
+
J_y \sin\theta_i\sin\theta_j
\Big].
\label{eq:H_nondiag_HMF}
\end{equation}
This model reduces to the isotropic HMF case for
\(
J_x=J_y=1
\)
and
\(
J_{xy}=0
\),
and to the diagonal anisotropic HMF case for
\(
J_{xy}=0
\)
with
\(
J_x\neq J_y
\).

Introducing the collective variables
\begin{equation}
m_x=\frac{1}{N}\sum_{i=1}^N \cos\theta_i,
\qquad
m_y=\frac{1}{N}\sum_{i=1}^N \sin\theta_i,
\label{eq:mx_my_nondiag}
\end{equation}
the interaction term can be rewritten as
\begin{equation}
\sum_{i,j}\cos\theta_i\cos\theta_j=N^2m_x^2,
\qquad
\sum_{i,j}\sin\theta_i\sin\theta_j=N^2m_y^2,
\end{equation}
and
\begin{equation}
\sum_{i,j}
\bigl(\cos\theta_i\sin\theta_j+\sin\theta_i\cos\theta_j\bigr)
=
2N^2m_xm_y.
\end{equation}
Therefore the potential takes the collective form
\begin{equation}
V
=
\frac{N}{2}
-
\frac{N}{2}
\Big(
J_x m_x^2+2J_{xy}m_xm_y+J_y m_y^2
\Big).
\label{eq:V_nondiag_HMF}
\end{equation}
Equivalently,
\begin{equation}
V
=
\frac{N}{2}
-
\frac{N}{2}\,\bm m^{\mathsf T}J\,\bm m,
\qquad
\bm m=
\begin{pmatrix}
m_x\\
m_y
\end{pmatrix},
\qquad
J=
\begin{pmatrix}
J_x & J_{xy}\\
J_{xy} & J_y
\end{pmatrix}.
\label{eq:Jmatrix_nondiag}
\end{equation}

This is precisely the type of non-diagonal collective sector anticipated by the general formalism. On the disordered branch, the relevant harmonic sector is still \(k=1\), so that
\begin{equation}
D=-\mathbb I_2,
\qquad
Q_\ast=\frac{1}{2}\mathbb I_2,
\qquad
v_0=\frac{1}{2}.
\label{eq:branch_data_nondiag}
\end{equation}
Using Eq.~\eqref{eq:C_general_explicit}, the collective curvature form becomes
\begin{equation}
\mathcal C(\varepsilon)
=
\frac{1}{c(\varepsilon)^2}
\left(
c(\varepsilon)J-\frac{1}{2}J^2
\right),
\qquad
c(\varepsilon)=2\varepsilon-1.
\label{eq:C_nondiag_compact}
\end{equation}
Explicitly,
\begin{equation}
\mathcal C(\varepsilon)
=
\frac{1}{(2\varepsilon-1)^2}
\begin{pmatrix}
(2\varepsilon-1)J_x-\frac{1}{2}(J_x^2+J_{xy}^2)
&
(2\varepsilon-1)J_{xy}-\frac{1}{2}J_{xy}(J_x+J_y)
\\[1mm]
(2\varepsilon-1)J_{xy}-\frac{1}{2}J_{xy}(J_x+J_y)
&
(2\varepsilon-1)J_y-\frac{1}{2}(J_y^2+J_{xy}^2)
\end{pmatrix}.
\label{eq:C_nondiag_explicit}
\end{equation}

The critical channels are no longer the bare directions \(m_x\) and \(m_y\), but the eigenvectors of \(J\), equivalently of \(\mathcal C(\varepsilon)\). The eigenvalues of \(J\) are
\begin{equation}
\lambda_\pm
=
\frac{J_x+J_y}{2}
\pm
\frac{1}{2}\sqrt{(J_x-J_y)^2+4J_{xy}^2},
\label{eq:lambda_pm_nondiag}
\end{equation}
and the corresponding curvature eigenvalues are
\begin{equation}
\mu_\pm(\varepsilon)
=
\frac{1}{c(\varepsilon)^2}
\left(
c(\varepsilon)\lambda_\pm-\frac{\lambda_\pm^2}{2}
\right)
=
\frac{\lambda_\pm}{c(\varepsilon)^2}
\left(
c(\varepsilon)-\frac{\lambda_\pm}{2}
\right).
\label{eq:mu_pm_nondiag}
\end{equation}
The geometric critical conditions are therefore
\begin{equation}
\mu_\pm(\varepsilon_{c,\pm})=0
\qquad \Longleftrightarrow \qquad
c(\varepsilon_{c,\pm})=\frac{\lambda_\pm}{2},
\label{eq:critical_cond_nondiag}
\end{equation}
which gives
\begin{equation}
\varepsilon_{c,\pm}
=
\frac{1}{2}+\frac{\lambda_\pm}{4}.
\label{eq:epsc_pm_nondiag}
\end{equation}
Substituting Eq.~\eqref{eq:lambda_pm_nondiag}, one finally obtains
\begin{equation}
\varepsilon_{c,\pm}
=
\frac{1}{2}
+
\frac{1}{8}
\left[
J_x+J_y
\pm
\sqrt{(J_x-J_y)^2+4J_{xy}^2}
\right].
\label{eq:epsc_pm_nondiag_explicit}
\end{equation}

This model makes the spectral content of the formalism fully explicit. When \(J_{xy}=0\), one recovers the diagonal anisotropic thresholds
\begin{equation}
\varepsilon_{c,x}=\frac{1}{2}+\frac{J_x}{4},
\qquad
\varepsilon_{c,y}=\frac{1}{2}+\frac{J_y}{4},
\label{eq:epsc_diag_limit}
\end{equation}
while for \(J_{xy}\neq 0\) the critical directions are rotated linear combinations of \(m_x\) and \(m_y\). The transition is then controlled not by a preassigned order parameter, but by the spectrum of the collective curvature form. This is the simplest explicit realization of a genuinely non-diagonal critical sector within the mean-field rotor class.

\subsection{A mixed first--second harmonic rotor model}
\label{subsec:mixed_harmonic_model}

A natural non-diagonal extension of the generalized HMF class is obtained by coupling the first and second harmonic sectors already at the level of the interaction. We consider the Hamiltonian
\begin{equation}
H
=
\sum_{i=1}^N \frac{p_i^2}{2}
+
\frac{1}{2N}\sum_{i,j=1}^N
\Big[
1
-
\Delta_1 \cos(\theta_i-\theta_j)
-
\Delta_2 \cos\bigl(2(\theta_i-\theta_j)\bigr)
\Big]
-
\frac{\gamma}{N}\sum_{i,j=1}^N \cos\theta_i \cos 2\theta_j ,
\label{eq:H_mixed_harmonics_realspace}
\end{equation}
where \(\Delta_1,\Delta_2\ge 0\) and \(\gamma\) controls the mixing between the \(k=1\) and \(k=2\) cosine sectors.

The first two terms are the natural extension of the GHMF interaction to the first and second harmonics, while the last term introduces an explicit quadratic mixing between the corresponding collective modes. Using
\begin{equation}
\cos(\theta_i-\theta_j)=\cos\theta_i\cos\theta_j+\sin\theta_i\sin\theta_j,
\label{eq:cos_diff_k1}
\end{equation}
and
\begin{equation}
\cos\bigl(2(\theta_i-\theta_j)\bigr)=\cos 2\theta_i\cos 2\theta_j+\sin 2\theta_i\sin 2\theta_j,
\label{eq:cos_diff_k2}
\end{equation}
we introduce the four collective variables
\begin{equation}
m_{c,1}=\frac{1}{N}\sum_{i=1}^N \cos\theta_i,
\qquad
m_{s,1}=\frac{1}{N}\sum_{i=1}^N \sin\theta_i,
\label{eq:mc1_ms1_def}
\end{equation}
\begin{equation}
m_{c,2}=\frac{1}{N}\sum_{i=1}^N \cos 2\theta_i,
\qquad
m_{s,2}=\frac{1}{N}\sum_{i=1}^N \sin 2\theta_i.
\label{eq:mc2_ms2_def}
\end{equation}
In terms of these quantities,
\begin{equation}
\sum_{i,j}\cos(\theta_i-\theta_j)=N^2(m_{c,1}^2+m_{s,1}^2),
\label{eq:sum_k1_collective}
\end{equation}
\begin{equation}
\sum_{i,j}\cos\bigl(2(\theta_i-\theta_j)\bigr)=N^2(m_{c,2}^2+m_{s,2}^2),
\label{eq:sum_k2_collective}
\end{equation}
and
\begin{equation}
\sum_{i,j}\cos\theta_i\cos 2\theta_j=N^2\,m_{c,1}m_{c,2}.
\label{eq:mixed_sum_collective}
\end{equation}
Therefore the potential can be rewritten as
\begin{equation}
V
=
\frac{N}{2}
-
\frac{N}{2}
\Big[
\Delta_1(m_{c,1}^2+m_{s,1}^2)
+
\Delta_2(m_{c,2}^2+m_{s,2}^2)
+
2\gamma\,m_{c,1}m_{c,2}
\Big].
\label{eq:V_mixed_collective}
\end{equation}

This is of the general quadratic collective form discussed in Sec.~\ref{sec:general_rotor_class}. Introducing
\begin{equation}
\bm m=
\begin{pmatrix}
m_{c,1}\\
m_{s,1}\\
m_{c,2}\\
m_{s,2}
\end{pmatrix},
\label{eq:mvec_mixed}
\end{equation}
the potential reads
\begin{equation}
V
=
\frac{N}{2}
-
\frac{N}{2}\,\bm m^{\mathsf T}J\,\bm m,
\label{eq:V_mixed_matrix}
\end{equation}
with coupling matrix
\begin{equation}
J=
\begin{pmatrix}
\Delta_1 & 0 & \gamma & 0\\
0 & \Delta_1 & 0 & 0\\
\gamma & 0 & \Delta_2 & 0\\
0 & 0 & 0 & \Delta_2
\end{pmatrix}.
\label{eq:J_mixed_model}
\end{equation}

The structure of Eq.~\eqref{eq:J_mixed_model} is already instructive. The sine sectors remain diagonal, while the cosine sectors of the first and second harmonics are explicitly coupled. This gives the simplest realization, within the rotor mean-field class, of a genuine non-diagonal collective sector involving different harmonics.

On the disordered branch, the harmonic closure matrices are
\begin{equation}
D=\mathrm{diag}(-1,-1,-4,-4),
\qquad
Q_\ast=\frac{1}{2}\,\mathrm{diag}(1,1,4,4),
\qquad
v_0=\frac{1}{2},
\label{eq:DQstar_mixed_model}
\end{equation}
and the kinetic background is
\begin{equation}
c(\varepsilon)=2(\varepsilon-v_0)=2\varepsilon-1.
\label{eq:c_eps_mixed_model}
\end{equation}
The collective curvature form is then obtained from the general expression
\begin{equation}
\mathcal C(\varepsilon)
=
\frac{1}{c(\varepsilon)^2}
\Bigl(
c(\varepsilon)\,\mathcal M-\mathcal B
\Bigr),
\label{eq:C_general_mixed_model}
\end{equation}
with
\begin{equation}
\mathcal M=-\frac{1}{2}(JD+DJ),
\qquad
\mathcal B=JQ_\ast J.
\label{eq:MB_defs_mixed_model}
\end{equation}

Since the mixed term acts only in the cosine subsector, it is convenient to analyze the four collective channels separately.

\paragraph{(i) The pure \(\bm{m_{s,1}}\) sector.}

The first sine harmonic is unaffected by the mixing. Its curvature coefficient is
\begin{equation}
\mathcal C_{s,1}(\varepsilon)
=
\frac{\Delta_1}{c(\varepsilon)^2}
\left(
c(\varepsilon)-\frac{\Delta_1}{2}
\right).
\label{eq:C_s1_mixed_model}
\end{equation}
The corresponding critical condition is
\begin{equation}
c(\varepsilon_{s,1})=\frac{\Delta_1}{2},
\label{eq:c_s1_mixed_model}
\end{equation}
hence
\begin{equation}
\varepsilon_{s,1}=\frac{1}{2}+\frac{\Delta_1}{4}.
\label{eq:eps_s1_mixed_model}
\end{equation}

\paragraph{(ii) The pure \(\bm{m_{s,2}}\) sector.}

The second sine harmonic is likewise diagonal. Its curvature coefficient is
\begin{equation}
\mathcal C_{s,2}(\varepsilon)
=
\frac{4\Delta_2}{c(\varepsilon)^2}
\left(
c(\varepsilon)-\frac{\Delta_2}{2}
\right),
\label{eq:C_s2_mixed_model}
\end{equation}
which yields
\begin{equation}
c(\varepsilon_{s,2})=\frac{\Delta_2}{2},
\label{eq:c_s2_mixed_model}
\end{equation}
and therefore
\begin{equation}
\varepsilon_{s,2}=\frac{1}{2}+\frac{\Delta_2}{4}.
\label{eq:eps_s2_mixed_model}
\end{equation}

\paragraph{(iii) The mixed cosine sector \(\bm{(m_{c,1},m_{c,2})}\).}

The nontrivial content of the model is entirely concentrated in the cosine subsector. Restricting to
\begin{equation}
\bm m_c=
\begin{pmatrix}
m_{c,1}\\
m_{c,2}
\end{pmatrix},
\label{eq:mcos_subsector}
\end{equation}
the coupling matrix is
\begin{equation}
J_c=
\begin{pmatrix}
\Delta_1 & \gamma\\
\gamma & \Delta_2
\end{pmatrix},
\qquad
D_c=
\begin{pmatrix}
-1 & 0\\
0 & -4
\end{pmatrix},
\qquad
Q_c=
\frac{1}{2}
\begin{pmatrix}
1 & 0\\
0 & 4
\end{pmatrix}.
\label{eq:JcDcQc_defs}
\end{equation}
From Eq.~\eqref{eq:MB_defs_mixed_model} one finds
\begin{equation}
\mathcal M_c
=
-\frac{1}{2}(J_cD_c+D_cJ_c)
=
\begin{pmatrix}
\Delta_1 & \dfrac{5}{2}\gamma\\[2mm]
\dfrac{5}{2}\gamma & 4\Delta_2
\end{pmatrix},
\label{eq:Mc_mixed_model}
\end{equation}
and
\begin{equation}
\mathcal B_c
=
J_cQ_cJ_c
=
\begin{pmatrix}
\dfrac{\Delta_1^2}{2}+2\gamma^2 &
\dfrac{\gamma\Delta_1}{2}+2\gamma\Delta_2\\[2mm]
\dfrac{\gamma\Delta_1}{2}+2\gamma\Delta_2 &
\dfrac{\gamma^2}{2}+2\Delta_2^2
\end{pmatrix}.
\label{eq:Bc_mixed_model}
\end{equation}
Therefore the collective curvature form in the mixed cosine sector is
\begin{equation}
\mathcal C_c(\varepsilon)
=
\frac{1}{c(\varepsilon)^2}
\begin{pmatrix}
c\Delta_1-\left(\dfrac{\Delta_1^2}{2}+2\gamma^2\right)
&
\dfrac{5}{2}c\gamma-\left(\dfrac{\gamma\Delta_1}{2}+2\gamma\Delta_2\right)
\\[3mm]
\dfrac{5}{2}c\gamma-\left(\dfrac{\gamma\Delta_1}{2}+2\gamma\Delta_2\right)
&
4c\Delta_2-\left(\dfrac{\gamma^2}{2}+2\Delta_2^2\right)
\end{pmatrix},
\label{eq:Cc_mixed_model}
\end{equation}
where \(c=c(\varepsilon)\).

The two critical energies associated with this sector are obtained from
\begin{equation}
\det \mathcal C_c(\varepsilon)=0.
\label{eq:detCc_zero}
\end{equation}
Since the global factor \(1/c(\varepsilon)^4\) is irrelevant for the zeros, one only needs the determinant of the numerator. A straightforward calculation gives
\begin{equation}
\left(4\Delta_1\Delta_2-\frac{25}{4}\gamma^2\right)c^2
-
2(\Delta_1+\Delta_2)(\Delta_1\Delta_2-\gamma^2)c
+
(\Delta_1\Delta_2-\gamma^2)^2
=0.
\label{eq:quadratic_c_mixed_model}
\end{equation}
The two solutions are
\begin{equation}
c_\pm
=
\frac{
2(\Delta_1\Delta_2-\gamma^2)
\left[
2(\Delta_1+\Delta_2)\pm\sqrt{4(\Delta_1-\Delta_2)^2+25\gamma^2}
\right]
}{
16\Delta_1\Delta_2-25\gamma^2
}.
\label{eq:cpm_mixed_model}
\end{equation}
Using Eq.~\eqref{eq:c_eps_mixed_model}, the corresponding critical energies are
\begin{equation}
\varepsilon_\pm
=
\frac{1}{2}
+
\frac{
(\Delta_1\Delta_2-\gamma^2)
\left[
2(\Delta_1+\Delta_2)\pm\sqrt{4(\Delta_1-\Delta_2)^2+25\gamma^2}
\right]
}{
16\Delta_1\Delta_2-25\gamma^2
}.
\label{eq:epspm_mixed_model}
\end{equation}

This example makes the role of the collective curvature form fully explicit. The sine sectors remain governed by the usual harmonic thresholds, while the cosine sectors of the first and second harmonics are mixed into two new critical channels selected by the spectrum of the \(2\times 2\) curvature block \(\mathcal C_c(\varepsilon)\). The transition is therefore no longer tied to a single harmonic order parameter, but to the spectral decomposition of the collective geometry itself.

\section{Critical energies from conventional statistical mechanics: canonical partition function, self-consistency equations, and Landau criticality}
\label{sec:canonical_benchmark}

The geometric critical energies derived in the previous sections admit an independent benchmark from the canonical solution of the same mean-field class. For Hamiltonians with quadratic collective potentials, the canonical problem can be reduced, in the thermodynamic limit, to a finite-dimensional variational principle. This yields the self-consistency equations for the order parameters and, in particular, the quadratic Landau form controlling the stability of the disordered branch. The resulting canonical thresholds provide the natural thermodynamic counterpart of the spectral-geometric critical condition.

\subsection{General mean-field rotor class}

We consider again the class of Hamiltonians
\begin{equation}
H(\theta,p)
=
\sum_{i=1}^N \frac{p_i^2}{2}
+
N v_0
-
\frac{N}{2}\,\bm m^{\mathsf T}J\,\bm m,
\label{eq:H_canonical_general}
\end{equation}
where
\begin{equation}
\bm m
=
\frac{1}{N}\sum_{i=1}^N \Phi(\theta_i),
\qquad
\Phi(\theta)=\bigl(\phi_1(\theta),\dots,\phi_r(\theta)\bigr)^{\mathsf T},
\label{eq:mPhi_canonical_general}
\end{equation}
and \(J=J^{\mathsf T}\) is a real symmetric matrix. Recall that the \(\phi_a\) are assumed to be bounded trigonometric modes, so the collective sector is finite-dimensional and the canonical reduction is well defined.

The canonical partition function reads
\begin{equation}
Z_N(\beta)
=
\int \prod_{i=1}^N dp_i\,d\theta_i\;
\exp\!\left[-\beta H(\theta,p)\right].
\label{eq:ZN_def}
\end{equation}
The Gaussian integration over the momenta gives
\begin{equation}
Z_N(\beta)
=
\left(\frac{2\pi}{\beta}\right)^{N/2}
e^{-\beta N v_0}
\int \prod_{i=1}^N d\theta_i\;
\exp\!\left[
\frac{\beta N}{2}\,\bm m^{\mathsf T}J\,\bm m
\right].
\label{eq:ZN_after_momenta}
\end{equation}

\subsection{Hubbard--Stratonovich representation and variational free energy}

Assume first that \(J\) is invertible on the relevant collective sector. Using a Hubbard--Stratonovich transformation, one may write
\begin{equation}
\exp\!\left[
\frac{\beta N}{2}\,\bm m^{\mathsf T}J\,\bm m
\right]
\propto
\int_{\mathbb R^r} d\bm x\;
\exp\!\left[
-\frac{\beta N}{2}\,\bm x^{\mathsf T}J^{-1}\bm x
+
\beta N\,\bm x^{\mathsf T}\bm m
\right].
\label{eq:HS_general}
\end{equation}
Since
\begin{equation}
N\,\bm x^{\mathsf T}\bm m
=
\sum_{i=1}^N \bm x^{\mathsf T}\Phi(\theta_i),
\label{eq:factorization_argument}
\end{equation}
the angular integrations factorize and one obtains
\begin{equation}
Z_N(\beta)
\propto
\left(\frac{2\pi}{\beta}\right)^{N/2}
e^{-\beta N v_0}
\int_{\mathbb R^r} d\bm x\;
\exp\!\left\{
-N
\left[
\frac{\beta}{2}\bm x^{\mathsf T}J^{-1}\bm x
-
\ln z_\theta(\beta,\bm x)
\right]
\right\},
\label{eq:ZN_HS_factorized}
\end{equation}
where
\begin{equation}
z_\theta(\beta,\bm x)
=
\int_{-\pi}^{\pi} d\theta\;
\exp\!\big[\beta\,\bm x^{\mathsf T}\Phi(\theta)\big].
\label{eq:ztheta_general}
\end{equation}

In the thermodynamic limit \(N\to\infty\), the integral is dominated by a saddle point. It is more convenient, however, to rewrite the result directly in terms of the collective variables \(\bm m\). The canonical free-energy density is then obtained from the variational functional
\begin{equation}
f(\beta;\bm m)
=
-\frac{1}{2\beta}\ln\frac{2\pi}{\beta}
+
v_0
+
\frac{1}{2}\,\bm m^{\mathsf T}J\,\bm m
-
\frac{1}{\beta}
\ln
\int_{-\pi}^{\pi}\frac{d\theta}{2\pi}\,
\exp\!\big[\beta\,(J\bm m)\!\cdot\!\Phi(\theta)\big].
\label{eq:f_variational_general}
\end{equation}
The physical free-energy density is therefore
\begin{equation}
f(\beta)=\inf_{\bm m} f(\beta;\bm m).
\label{eq:f_inf_general}
\end{equation}

Equation~\eqref{eq:f_variational_general} is the exact canonical solution of the mean-field problem in variational form. It is implicit, in the sense that the minimizing order parameters generally have to be determined from self-consistency equations, but it reduces the full thermodynamic problem to a finite-dimensional one.

\subsection{General self-consistency equations}

Stationarity of Eq.~\eqref{eq:f_variational_general} with respect to \(m_a\) gives
\begin{equation}
m_a
=
\frac{
\displaystyle
\int_{-\pi}^{\pi}\frac{d\theta}{2\pi}\,
\phi_a(\theta)\,
\exp\!\big[\beta\,(J\bm m)\!\cdot\!\Phi(\theta)\big]
}{
\displaystyle
\int_{-\pi}^{\pi}\frac{d\theta}{2\pi}\,
\exp\!\big[\beta\,(J\bm m)\!\cdot\!\Phi(\theta)\big]
},
\qquad
a=1,\dots,r\,,
\label{eq:self_consistency_components}
\end{equation}
that, in vector form, reads:
\begin{equation}
\bm m
=
\frac{
\displaystyle
\int_{-\pi}^{\pi}\frac{d\theta}{2\pi}\,
\Phi(\theta)\,
\exp\!\big[\beta\,(J\bm m)\!\cdot\!\Phi(\theta)\big]
}{
\displaystyle
\int_{-\pi}^{\pi}\frac{d\theta}{2\pi}\,
\exp\!\big[\beta\,(J\bm m)\!\cdot\!\Phi(\theta)\big]
}.
\label{eq:self_consistency_vector}
\end{equation}
These equations determine the canonical equilibrium branch \(\bm m_\ast(\beta)\). Once \(\bm m_\ast(\beta)\) is known, the corresponding canonical energy density is
\begin{equation}
\varepsilon(\beta)
=
\frac{1}{2\beta}
+
v_0
-
\frac{1}{2}\,\bm m_\ast(\beta)^{\mathsf T}J\,\bm m_\ast(\beta).
\label{eq:epsilon_beta_general}
\end{equation}

\subsection{Landau expansion around the disordered branch}

The canonical critical thresholds are determined by the quadratic stability of the disordered branch \(\bm m=\bm 0\). To extract them, we expand Eq.~\eqref{eq:f_variational_general} for small \(\bm m\).

Define the branch covariance matrix
\begin{equation}
Q_0
=
\int_{-\pi}^{\pi}\frac{d\theta}{2\pi}\,
\Phi(\theta)\Phi(\theta)^{\mathsf T}.
\label{eq:Q0_general}
\end{equation}
Since the disordered branch satisfies
\(
\int d\theta\,\Phi(\theta)=0
\)
for the trigonometric families of interest, one has
\begin{equation}
\ln
\int_{-\pi}^{\pi}\frac{d\theta}{2\pi}\,
\exp\!\big[\beta\,(J\bm m)\!\cdot\!\Phi(\theta)\big]
=
\frac{\beta^2}{2}\,
(J\bm m)^{\mathsf T}Q_0(J\bm m)
+
O(\|\bm m\|^4).
\label{eq:log_expansion_general}
\end{equation}
Substituting into Eq.~\eqref{eq:f_variational_general}, one obtains the Landau expansion
\begin{equation}
f(\beta;\bm m)
=
f_0(\beta)
+
\frac{1}{2}\,
\bm m^{\mathsf T}
\Bigl(
J-\beta JQ_0J
\Bigr)
\bm m
+
O(\|\bm m\|^4),
\label{eq:Landau_general}
\end{equation}
where
\begin{equation}
f_0(\beta)
=
-\frac{1}{2\beta}\ln\frac{2\pi}{\beta}+v_0
\label{eq:f0_general}
\end{equation}
is the free-energy density of the disordered phase. The quadratic form
\begin{equation}
A(\beta):=J-\beta JQ_0J
\label{eq:A_beta_general}
\end{equation}
controls the canonical stability of the disordered branch.

\subsection{General canonical critical condition}

A canonical critical threshold is reached when the quadratic form in Eq.~\eqref{eq:Landau_general} becomes marginal. Therefore the general canonical critical condition is
\begin{equation}
\det A(\beta_c)=0,
\label{eq:detA_zero_general}
\end{equation}
that is,
\begin{equation}
\det\!\bigl(J-\beta_c JQ_0J\bigr)=0.
\label{eq:detJbetaQ_general}
\end{equation}
If \(J\) is invertible on the relevant sector, this is equivalent to
\begin{equation}
\det\!\bigl(\mathbb I-\beta_c Q_0J\bigr)=0.
\label{eq:detIbetaQJ_general}
\end{equation}
Hence the critical inverse temperatures are determined by the spectrum of \(Q_0J\):
\begin{equation}
\beta_{c,\alpha}=\frac{1}{\mu_\alpha},
\qquad
\mu_\alpha\in\text{Spec}(Q_0J).
\label{eq:beta_c_general_spectrum}
\end{equation}
On the disordered branch, where
\begin{equation}
\varepsilon=\frac{1}{2\beta}+v_0,
\label{eq:epsilon_disordered_general}
\end{equation}
the corresponding critical energies are
\begin{equation}
\varepsilon_{c,\alpha}
=
v_0+\frac{\mu_\alpha}{2}.
\label{eq:epsilon_c_general_spectrum}
\end{equation}

Equations~\eqref{eq:beta_c_general_spectrum}--\eqref{eq:epsilon_c_general_spectrum} provide the general canonical benchmark for the geometric spectrum derived from the collective curvature form.

\subsection{Recovery of standard examples}

We now specialize the general formula to the models discussed above.

\paragraph{(i) Isotropic HMF model.}

For
\begin{equation}
\Phi(\theta)=
\begin{pmatrix}
\cos\theta\\
\sin\theta
\end{pmatrix},
\qquad
J=\mathbb I_2,
\qquad
v_0=\frac{1}{2},
\label{eq:HMF_canonical_data}
\end{equation}
one has
\begin{equation}
Q_0=\frac{1}{2}\mathbb I_2.
\label{eq:HMF_Q0}
\end{equation}
Therefore
\begin{equation}
Q_0J=\frac{1}{2}\mathbb I_2,
\qquad
\beta_c=2,
\qquad
\varepsilon_c=\frac{1}{2}+\frac{1}{4}=\frac{3}{4},
\label{eq:HMF_canonical_crit}
\end{equation}
which reproduces the standard HMF threshold.

\paragraph{(ii) Anisotropic HMF model.}

For
\begin{equation}
J=
\begin{pmatrix}
1+D_a & 0\\
0 & 1
\end{pmatrix},
\qquad
Q_0=\frac{1}{2}\mathbb I_2,
\qquad
v_0=\frac{1}{2},
\label{eq:aniso_canonical_data}
\end{equation}
the eigenvalues of \(Q_0J\) are
\begin{equation}
\mu_x=\frac{1+D_a}{2},
\qquad
\mu_y=\frac{1}{2}.
\label{eq:aniso_mu}
\end{equation}
Thus
\begin{equation}
\beta_{c,x}=\frac{2}{1+D_a},
\qquad
\beta_{c,y}=2,
\label{eq:aniso_betac}
\end{equation}
and
\begin{equation}
\varepsilon_{c,x}=\frac{1}{2}+\frac{1+D_a}{4}=\frac{3+D_a}{4},
\qquad
\varepsilon_{c,y}=\frac{3}{4}.
\label{eq:aniso_epsc}
\end{equation}
The dominant canonical threshold coincides with the geometric one.

\paragraph{(iii) GHMF model.}

For the standard GHMF model, the collective basis is
\begin{equation}
\Phi(\theta)=
\begin{pmatrix}
\cos\theta\\
\sin\theta\\
\cos2\theta\\
\sin2\theta
\end{pmatrix},
\qquad
J=
\text{diag}(\Delta,\Delta,1-\Delta,1-\Delta),
\qquad
v_0=\frac{1}{2}.
\label{eq:GHMF_canonical_data}
\end{equation}
On the disordered branch,
\begin{equation}
Q_0=\frac{1}{2}\mathbb I_4.
\label{eq:GHMF_Q0}
\end{equation}
Hence the nonzero distinct eigenvalues of \(Q_0J\) are
\begin{equation}
\mu_1=\frac{\Delta}{2},
\qquad
\mu_2=\frac{1-\Delta}{2},
\label{eq:GHMF_mu}
\end{equation}
which give
\begin{equation}
\varepsilon_c^{(1)}=\frac{1}{2}+\frac{\Delta}{4},
\qquad
\varepsilon_c^{(2)}=\frac{1}{2}+\frac{1-\Delta}{4}
=
\frac{3-\Delta}{4}.
\label{eq:GHMF_epsc_canonical}
\end{equation}
These coincide with the critical energies obtained from the collective curvature form.

\paragraph{(iv) Mixed first--second harmonic model.}

For the mixed model of Sec.~\ref{subsec:mixed_harmonic_model},
\begin{equation}
\Phi(\theta)=
\begin{pmatrix}
\cos\theta\\
\sin\theta\\
\cos2\theta\\
\sin2\theta
\end{pmatrix},
\qquad
J=
\begin{pmatrix}
\Delta_1 & 0 & \gamma & 0\\
0 & \Delta_1 & 0 & 0\\
\gamma & 0 & \Delta_2 & 0\\
0 & 0 & 0 & \Delta_2
\end{pmatrix},
\qquad
v_0=\frac{1}{2},
\label{eq:mixed_canonical_data}
\end{equation}
while again
\begin{equation}
Q_0=\frac{1}{2}\mathbb I_4.
\label{eq:mixed_Q0}
\end{equation}
Therefore
\begin{equation}
Q_0J=\frac{1}{2}J.
\label{eq:mixed_Q0J}
\end{equation}
Two eigenvalues are immediately read off from the uncoupled sine channels:
\begin{equation}
\mu_{s,1}=\frac{\Delta_1}{2},
\qquad
\mu_{s,2}=\frac{\Delta_2}{2},
\label{eq:mixed_mu_sine}
\end{equation}
yielding
\begin{equation}
\varepsilon_{s,1}=\frac{1}{2}+\frac{\Delta_1}{4},
\qquad
\varepsilon_{s,2}=\frac{1}{2}+\frac{\Delta_2}{4}.
\label{eq:mixed_epsc_sine}
\end{equation}
The mixed cosine block is
\begin{equation}
\frac{1}{2}
\begin{pmatrix}
\Delta_1 & \gamma\\
\gamma & \Delta_2
\end{pmatrix},
\label{eq:mixed_cosine_block}
\end{equation}
whose eigenvalues are
\begin{equation}
\mu_\pm
=
\frac{1}{4}
\left[
\Delta_1+\Delta_2
\pm
\sqrt{(\Delta_1-\Delta_2)^2+4\gamma^2}
\right].
\label{eq:mixed_mu_pm}
\end{equation}
Thus
\begin{equation}
\varepsilon_{c,\pm}
=
\frac{1}{2}
+
\frac{1}{8}
\left[
\Delta_1+\Delta_2
\pm
\sqrt{(\Delta_1-\Delta_2)^2+4\gamma^2}
\right].
\label{eq:mixed_epsc_canonical}
\end{equation}
These are the canonical critical energies associated with the mixed first--second harmonic sector.

\medskip
The canonical results above provide the benchmark required to interpret the geometric spectrum. For the whole class \eqref{eq:H_canonical_general}, the canonical stability problem of the disordered branch is governed by the spectrum of \(Q_0J\), while the geometric critical condition derived earlier is governed by the spectrum of the collective curvature form \(\mathcal C(\varepsilon)\). In the examples treated here, the two constructions select the same critical thresholds. This agreement is precisely the expected consistency check: the geometric spectrum extracted from \(\text{Tr} W_\xi\) reproduces the correct thermodynamic critical energies of the corresponding mean-field models.

\section{Discussion and Conclusions}

This work shows that the onset of criticality is governed by a loss of geometric rigidity of the constant-energy shell \(\Sigma_E\) along specific collective directions. The key point is that this mechanism is not model-specific in form. Near a given reference phase branch, the local extrinsic geometry of \(\Sigma_E\), as encoded by the trace of the Weingarten operator, organizes into a finite-dimensional collective sector and admits a universal quadratic expansion in the order parameters:
\begin{equation}
\frac{\text{Tr} W_\xi}{N}
=
\frac{1}{c(\varepsilon)}
+
\bm m^{\mathsf T}\,\mathcal C(\varepsilon)\,\bm m
+
O(\|\bm m\|^3),\qquad \text{with}\quad \mathcal{C}(\epsilon)\bm{v}_\alpha=\lambda_\alpha(\epsilon)\bm{v}_\alpha.
\label{eq:discussion_expansion}
\end{equation}
The corresponding collective curvature form \(\mathcal C(\varepsilon)\) selects the relevant critical channels through its eigenvectors $\bm{v}_\alpha$, while the critical energies are determined by the vanishing of the corresponding eigenvalues, $\lambda_\alpha(\epsilon_c)=0$ or $\det\mathcal{C}(\epsilon_c)=0$. In this sense, the transition is generated by a precise geometric instability.

This is, in our view, the central conceptual contribution of the paper. For the class of systems studied here, criticality is not only a thermodynamic phenomenon to be detected a posteriori, but the macroscopic manifestation of an underlying geometric reorganization of the energy shell. The result therefore identifies a structural level at which the transition is already encoded before it is read in terms of singularities. Thermodynamics remains fully valid, but its role changes: it becomes the language in which a deeper geometric mechanism becomes observable.

A second important aspect is the universality of this mechanism within the class considered. The result does not depend on the detailed form of each Hamiltonian separately, but on the fact that the interaction closes on a finite set of collective trigonometric modes. Within this class, the onset of criticality is always governed by the geometric object, namely the scalar curvature, $\text{Tr}W_{\bm\xi}$. 

The significance of this point is especially clear in the context of long-range and mean-field systems. These models are not only analytically tractable but also conceptually demanding because nonadditivity and ensemble inequivalence challenge the automatic primacy of canonical descriptions. In such regimes, the microcanonical setting is not merely an alternative formulation: it is the natural place where the geometry of the energy shell becomes directly relevant. Showing that the present framework remains predictive precisely in this setting is therefore a substantive test of its scope. The result suggests that geometric microcanonical methods are not restricted to benign cases where standard equilibrium approaches trivially agree but remain meaningful where the structure of the transition is most delicate.

At the same time, the scope of the present criterion should be stated clearly. The mechanism identified here is local with respect to the phase branch around which the geometric expansion is performed. This is not a limitation of principle but an intrinsic feature of systems with multiple competing phases. Different transitions may originate from different backgrounds and therefore require distinct local expansions. The generalized HMF model illustrates this point clearly: the expansion around the disordered branch captures the \(P\to F\) and \(P\to N\) instabilities, whereas the \(N\to F\) transition requires a second expansion around the nematic branch. The mechanism itself does not change; what changes is the branch on which it becomes active.

A further distinction must also be emphasized. Identifying the mechanism of criticality is not the same as classifying the order of the transition. The present work addresses the first problem: it tells us which collective geometric direction loses rigidity and at which energy this happens. The second problem requires an additional thermodynamic analysis of entropy derivatives and response functions, especially at finite \(N\). In this sense, the present framework is naturally complementary to approaches such as microcanonical inflection-point analysis: geometry identifies the structural origin of the transition, while the order and hierarchy of the transition must still be determined through the corresponding thermodynamic observables.

Taken together, these results show that, for a broad and natural class of mean-field rotor Hamiltonians, phase transitions can be understood as the macroscopic manifestation of a microscopic geometric mechanism: a loss of rigidity of the constant-energy shell along distinguished collective directions. This does not replace thermodynamics, but identifies the structural level at which criticality is organized before becoming thermodynamically visible.

\section*{Acknowledgement}
I dedicate this work to my wife, Sara, because through the immutable strength of her presence, in the cruel clutch of circumstance I shall not wince, nor shall I cry aloud; and however merciless the sentence may be, my dignity shall remain unconquered.

\appendix

\section{Collective expansion of \texorpdfstring{$\Delta H$}{Delta H}, \texorpdfstring{$\|\nabla H\|^2$}{||grad H||^2}, and \texorpdfstring{$\mathrm{Tr}\,W_{\bm\xi}$}{Tr Wxi}}
\label{app:collective_expansion}

In this Appendix we derive the collective expansion of
\(
\Delta H
\),
\(
\|\nabla H\|^2
\),
and
\(
\mathrm{Tr}\,W_{\bm\xi}
\)
for mean-field Hamiltonians of the form discussed in the main text.

\subsection{Collective variables and reference branch}

Let
\begin{equation}
\Phi(\theta)
=
\bigl(\phi_1(\theta),\dots,\phi_r(\theta)\bigr)^{\mathsf T},
\end{equation}
where the functions \(\phi_a(\theta)\) form a finite trigonometric family closed under differentiation up to a constant matrix \(D\), namely
\begin{equation}
\Phi''(\theta)=-D\,\Phi(\theta).
\label{eq:Phi_second_derivative_app}
\end{equation}
The collective coordinates are
\begin{equation}
\bm m
=
\frac{1}{N}\sum_{i=1}^N \Phi(\theta_i)\in\mathbb R^r.
\label{eq:m_def_app}
\end{equation}

To perform a controlled expansion, we consider a family of microscopic configurations
\(
\Gamma_N(\bm m)=\{(\theta_i,p_i)\}_{i=1}^N
\)
parametrized by \(\bm m\) in a neighborhood of a reference value
\(
\bm m_\ast
\).
We call such a family a \emph{branch} if, for every \(\bm m\) in that neighborhood, the corresponding configuration satisfies
\eqref{eq:m_def_app}
and if the empirical averages of the relevant one-body observables admit deterministic large-\(N\) limits along the family.

In particular, we define the microscopic derivative covariance matrix
\begin{equation}
Q_N(\bm m)
:=
\frac{1}{N}\sum_{i=1}^N
\Phi'(\theta_i)\Phi'(\theta_i)^{\mathsf T}.
\label{eq:QN_def_app}
\end{equation}
Along the chosen branch, we assume that \(Q_N(\bm m)\) converges, for \(N\to\infty\), to a smooth matrix-valued function
\begin{equation}
Q(\bm m)
=
\lim_{N\to\infty} Q_N(\bm m).
\label{eq:Q_limit_branch_app}
\end{equation}
The \emph{reference value} of the derivative covariance matrix is then
\begin{equation}
Q_\ast := Q(\bm m_\ast).
\label{eq:Qstar_def_app}
\end{equation}
When the expansion is performed around the disordered branch, one simply has
\(
\bm m_\ast=\bm 0
\),
so that
\(
Q_\ast=Q(\bm 0)
\).

Thus, \(Q_\ast\) is the value, at the reference point of the branch, of the large-\(N\) limit of the matrix \(Q_N(\bm m)\).

\subsection{Exact expression for \texorpdfstring{$\Delta H$}{Delta H}}
\label{app:expansion-deltaH}

From the exact formula for the second derivatives,
\begin{equation}
\frac{\partial^2 H}{\partial \theta_i^2}
=
-
\frac{1}{N}\Phi'(\theta_i)^{\mathsf T}J\,\Phi'(\theta_i)
-
(J\bm m)^{\mathsf T}\Phi''(\theta_i),
\label{eq:second_diag_app}
\end{equation}
and using \eqref{eq:Phi_second_derivative_app}, one obtains
\begin{equation}
\frac{\partial^2 H}{\partial \theta_i^2}
=
-
\frac{1}{N}\Phi'(\theta_i)^{\mathsf T}J\,\Phi'(\theta_i)
+
(J\bm m)^{\mathsf T}D\,\Phi(\theta_i).
\label{eq:second_diag_withD_app}
\end{equation}
Summing over \(i\) gives
\begin{equation}
\Delta H
=
\sum_{i=1}^N \frac{\partial^2 H}{\partial p_i^2}+\sum_{i=1}^N \frac{\partial^2 H}{\partial \theta_i^2}
=
N
-
\frac{1}{N}\sum_{i=1}^N \Phi'(\theta_i)^{\mathsf T}J\,\Phi'(\theta_i)
+
N\,(J\bm m)^{\mathsf T}D\,\bm m.
\label{eq:DeltaH_exact_app}
\end{equation}
Since
\begin{equation}
\frac{1}{N}\sum_{i=1}^N \Phi'(\theta_i)^{\mathsf T}J\,\Phi'(\theta_i)
=
\mathrm{Tr}\!\bigl(JQ_N(\bm m)\bigr),
\label{eq:trace_identity_app}
\end{equation}
we may rewrite \eqref{eq:DeltaH_exact_app} as
\begin{equation}
\frac{\Delta H}{N}
=
1
-
\frac{1}{N}\,\mathrm{Tr}\!\bigl(JQ_N(\bm m)\bigr)
+
\bm m^{\mathsf T}JD\,\bm m.
\label{eq:DeltaH_overN_exact_app}
\end{equation}

\subsection{Expansion around the reference branch}
\label{app:expansion-deltaH-branch}
Assume now that the branch map \(Q(\bm m)\) is differentiable at \(\bm m_\ast\). Then, for \(\bm m\) close to \(\bm m_\ast\),
\begin{equation}
Q(\bm m)
=
Q_\ast
+
\delta Q(\bm m),
\qquad
\delta Q(\bm m)=O(\|\bm m-\bm m_\ast\|).
\label{eq:Q_expand_app}
\end{equation}
If the expansion is performed around \(\bm m_\ast=\bm 0\), this becomes simply
\begin{equation}
Q(\bm m)=Q_\ast+O(\|\bm m\|).
\label{eq:Q_expand_zero_app}
\end{equation}

Using
\(
Q_N(\bm m)=Q(\bm m)+o(1)
\)
for \(N\to\infty\), Eq.~\eqref{eq:DeltaH_overN_exact_app} yields
\begin{equation}
\frac{\Delta H}{N}
=
1
-
\frac{1}{N}\,\mathrm{Tr}(JQ_\ast)
+
\bm m^{\mathsf T}JD\,\bm m
+
O(\|\bm m\|^3)
+
O(N^{-1}).
\label{eq:DeltaH_preM_app}
\end{equation}
The term \(\mathrm{Tr}(JQ_\ast)/N\) is \(O(N^{-1})\) and therefore subleading in the thermodynamic limit.

It is convenient to rewrite the quadratic form through its symmetric part. Since for any matrix \(A\),
\(
\bm m^{\mathsf T}A\bm m
=
\bm m^{\mathsf T}\frac12(A+A^{\mathsf T})\bm m
\),
we define
\begin{equation}
\mathcal M
:=
\frac12\bigl(JD+D^{\mathsf T}J\bigr),
\label{eq:M_def_app}
\end{equation}
so that
\begin{equation}
\bm m^{\mathsf T}JD\,\bm m
=
\bm m^{\mathsf T}\mathcal M\,\bm m.
\label{eq:JD_symmetrized_app}
\end{equation}
Therefore
\begin{equation}
\frac{\Delta H}{N}
=
1
+
\bm m^{\mathsf T}\mathcal M\,\bm m
+
O(\|\bm m\|^3)
+
O(N^{-1}).
\label{eq:DeltaH_final_app}
\end{equation}

For the harmonic family, \(D=D^{\mathsf T}\), hence
\begin{equation}
\mathcal M=\frac12(JD+DJ).
\label{eq:M_harmonic_app}
\end{equation}
If in addition \(J\) and \(D\) commute, then
\begin{equation}
\mathcal M=JD.
\label{eq:M_commuting_app}
\end{equation}

\subsection{Exact expression for \texorpdfstring{$\|\nabla H\|^2$}{||grad H||^2}}
\label{app:expansion-G}

From the first derivatives,
\begin{equation}
\frac{\partial H}{\partial p_i}=p_i,
\qquad
\frac{\partial H}{\partial \theta_i}
=
(J\bm m)^{\mathsf T}\Phi'(\theta_i),
\label{eq:first_derivatives_app}
\end{equation}
the squared norm of the gradient is
\begin{equation}
\|\nabla H\|^2
=
\sum_{i=1}^N p_i^2
+
\sum_{i=1}^N
\Bigl[(J\bm m)^{\mathsf T}\Phi'(\theta_i)\Bigr]^2.
\label{eq:gradnorm_exact_app}
\end{equation}
Define
\begin{equation}
K_N:=\sum_{i=1}^N p_i^2,
\qquad
c_N:=\frac{K_N}{N}.
\label{eq:KN_cN_def_app}
\end{equation}
Then
\begin{equation}
\sum_{i=1}^N
\Bigl[(J\bm m)^{\mathsf T}\Phi'(\theta_i)\Bigr]^2
=
N\,\bm m^{\mathsf T}JQ_N(\bm m)J\,\bm m,
\label{eq:quadratic_Q_app}
\end{equation}
hence
\begin{equation}
\|\nabla H\|^2
=
K_N
+
N\,\bm m^{\mathsf T}JQ_N(\bm m)J\,\bm m.
\label{eq:gradnorm_Q_exact_app}
\end{equation}
Dividing by \(N\),
\begin{equation}
G_N
:=
\frac{\|\nabla H\|^2}{N}
=
c_N
+
\bm m^{\mathsf T}JQ_N(\bm m)J\,\bm m.
\label{eq:GN_def_app}
\end{equation}

Assuming
\(
c_N\to c
\)
and
\(
Q_N(\bm m)\to Q(\bm m)
\),
we obtain near the reference branch
\begin{equation}
G
:=
\lim_{N\to\infty}G_N
=
c
+
\bm m^{\mathsf T}\mathcal B\,\bm m
+
O(\|\bm m\|^3),
\label{eq:G_final_app}
\end{equation}
with
\begin{equation}
\mathcal B:=JQ_\ast J.
\label{eq:B_def_app}
\end{equation}

\subsection{Expansion of \texorpdfstring{$\mathrm{Tr}\,W_{\bm\xi}/N$}{Tr Wxi/N}}
\label{app:expansion-TrW}

At leading order in \(N\), the Hessian-contraction term entering \(\mathrm{Tr}\,W_{\bm\xi}\) is subleading, so that
\begin{equation}
\frac{\mathrm{Tr}\,W_{\bm\xi}}{N}
=
\frac{\Delta H/N}{G}
+
O(N^{-1}).
\label{eq:TrW_reduction_app}
\end{equation}
Substituting \eqref{eq:DeltaH_final_app} and \eqref{eq:G_final_app},
\begin{equation}
\frac{\mathrm{Tr}\,W_{\bm\xi}}{N}
=
\frac{1+\bm m^{\mathsf T}\mathcal M\bm m+O(\|\bm m\|^3)+O(N^{-1})}
{c+\bm m^{\mathsf T}\mathcal B\bm m+O(\|\bm m\|^3)}.
\label{eq:TrW_ratio_app}
\end{equation}
Using the expansion
\begin{equation}
\frac{1+a}{c+b}
=
\frac1c+\frac{a}{c}-\frac{b}{c^2}+O(a^2,ab,b^2),
\label{eq:ratio_expand_app}
\end{equation}
with
\(
a=\bm m^{\mathsf T}\mathcal M\bm m
\)
and
\(
b=\bm m^{\mathsf T}\mathcal B\bm m
\),
we find
\begin{equation}
\frac{\mathrm{Tr}\,W_{\bm\xi}}{N}
=
\frac{1}{c}
+
\bm m^{\mathsf T}\mathcal C(\varepsilon)\,\bm m
+
O(\|\bm m\|^3)
+
O(N^{-1}),
\label{eq:TrW_final_app}
\end{equation}
where
\begin{equation}
\mathcal C(\varepsilon)
=
\frac{1}{c^2}\bigl(c\,\mathcal M-\mathcal B\bigr).
\label{eq:C_def_app}
\end{equation}
Using the explicit expressions of \(\mathcal M\) and \(\mathcal B\), this becomes
\begin{equation}
\mathcal C(\varepsilon)
=
\frac{1}{c^2}
\left[
\frac{c}{2}\bigl(JD+D^{\mathsf T}J\bigr)-JQ_\ast J
\right].
\label{eq:C_explicit_app}
\end{equation}

\subsection{Subleading character of the Hessian-contraction term}
\label{app:subleading-hessian-force}

We justify the approximation used in Eq.~\eqref{eq:TrW_reduction_sec4} by showing that the second term in the exact expression
\begin{equation}
\mathrm{Tr}\,W_{\boldsymbol{\xi}} =
\frac{\Delta H}{\|\nabla H\|^2}
- 2\,\frac{\nabla H^{\top}(\mathrm{Hess}\,H)\nabla H}{\|\nabla H\|^4}
\label{eq:TrW_exact}
\end{equation}
is negligible at leading order in the thermodynamic limit and at quadratic order
in the collective modes $\boldsymbol{m}$. The argument parallels the block decomposition used in Ref.~\cite{di2025geometric} for the 1D XY mean-field model, and we generalize it here to the full class of mean-field rotor Hamiltonians.

\paragraph{Exact expression for the contraction.}
From the gradient
\begin{equation}
\nabla H = \bigl(p_i,\;-(J\boldsymbol{m})^{\top}\Phi'(\theta_i)\bigr)^{\top}
\end{equation}
and the Hessian
\begin{equation}
\frac{\partial^2 H}{\partial\theta_i\partial\theta_j}
= -\frac{1}{N}\Phi'(\theta_i)^{\top}J\Phi'(\theta_j)
  - \delta_{ij}(J\boldsymbol{m})^{\top}\Phi''(\theta_i),
\qquad
\frac{\partial^2 H}{\partial p_i\partial p_j} = \delta_{ij},
\end{equation}
a direct computation gives
\begin{equation}
\begin{split}
\nabla H^{\top}(\mathrm{Hess}\,H)\nabla H
= \underbrace{\sum_{i=1}^{N} p_i^2}_{=:\,K}
&+ \underbrace{\sum_{i=1}^{N}
  \bigl[(J\boldsymbol{m})^{\top}\Phi'(\theta_i)\bigr]^2
  (J\boldsymbol{m})^{\top}\Phi''(\theta_i)}_{\mathrm{(I)}}
\\
&- \underbrace{\frac{1}{N}\sum_{i,j=1}^{N}
  \bigl[(J\boldsymbol{m})^{\top}\Phi'(\theta_i)\bigr]
  \bigl[\Phi'(\theta_i)^{\top}J\Phi'(\theta_j)\bigr]
  \bigl[(J\boldsymbol{m})^{\top}\Phi'(\theta_j)\bigr]}_{\mathrm{(II)}}.
\end{split}
\label{eq:hess_contr}
\end{equation}
The full contraction therefore splits into three blocks: a kinetic block $K$, a diagonal angular block~(I), and an interaction block~(II). We estimate each in turn.

\paragraph{Block K: kinetic term.}
From Eq.~\eqref{eq:GN_def_app}, $\|\nabla H\|^4 = N^2 G^2$ with
$G = c + O(\|\boldsymbol{m}\|^2)$, so
\begin{equation}
\frac{2K}{\|\nabla H\|^4}
= \frac{2Nc}{N^2 c^2\bigl(1+O(\|\boldsymbol{m}\|^2)\bigr)}
= \frac{2}{Nc}\bigl(1+O(\|\boldsymbol{m}\|^2)\bigr)
= O(N^{-1}).
\end{equation}
This term is suppressed by one power of $N$ relative to the leading
$O(1)$ background $1/c(\varepsilon)$ in $\mathrm{Tr}\,W_{\boldsymbol{\xi}}/N$.

\paragraph{Block (I): diagonal angular term.}
Using the closure relation $\Phi''(\theta) = -D\Phi(\theta)$, block (I) becomes
\begin{equation}
\mathrm{(I)}
= -\sum_{i=1}^{N}
  \bigl[(J\boldsymbol{m})^{\top}\Phi'(\theta_i)\bigr]^2\,
  (J\boldsymbol{m})^{\top}D\,\Phi(\theta_i).
\end{equation}
Each factor $(J\boldsymbol{m})^{\top}\Phi'(\theta_i)$ is $O(\|\boldsymbol{m}\|)$
since $\Phi'$ is bounded and $J\boldsymbol{m} = O(\|\boldsymbol{m}\|)$.
Similarly, $(J\boldsymbol{m})^{\top}D\,\Phi(\theta_i) = O(\|\boldsymbol{m}\|)$.
Therefore each summand is $O(\|\boldsymbol{m}\|^3)$, and summing over $N$ sites:
\begin{equation}
\mathrm{(I)} = O\bigl(N\|\boldsymbol{m}\|^3\bigr).
\end{equation}
Its contribution to the second term of~\eqref{eq:TrW_exact} is
\begin{equation}
\frac{2\,|\mathrm{(I)}|}{\|\nabla H\|^4}
= \frac{O(N\|\boldsymbol{m}\|^3)}{O(N^2)}
= O\!\left(\frac{\|\boldsymbol{m}\|^3}{N}\right).
\end{equation}
This is suppressed \emph{both} by $N^{-1}$ and by one additional power of
$\|\boldsymbol{m}\|$ relative to the quadratic collective correction
$\boldsymbol{m}^{\top}\mathcal{Q}(\varepsilon)\boldsymbol{m} = O(\|\boldsymbol{m}\|^2)$
in $\mathrm{Tr}\,W_{\boldsymbol{\xi}}/N$.

\paragraph{Block (II): interaction term.}
Block~(II) can be rewritten using the empirical covariance matrix
$Q_N(\boldsymbol{m}) = N^{-1}\sum_i \Phi'(\theta_i)\Phi'(\theta_i)^{\top}$ as
\begin{equation}
\mathrm{(II)}
= \frac{1}{N}\sum_{i,j=1}^{N}
  \bigl[(J\boldsymbol{m})^{\top}\Phi'(\theta_i)\bigr]
  \bigl[\Phi'(\theta_i)^{\top}J\Phi'(\theta_j)\bigr]
  \bigl[(J\boldsymbol{m})^{\top}\Phi'(\theta_j)\bigr]
= N\,\boldsymbol{m}^{\top}J\,Q_N J\,Q_N J\,\boldsymbol{m}
\cdot O(1).
\end{equation}
More precisely, writing out the indices explicitly:
\begin{align}
\mathrm{(II)}
&= \frac{1}{N}\sum_{i,j}
   (J\boldsymbol{m})^{\top}\Phi'(\theta_i)\,
   \Phi'(\theta_i)^{\top}J\Phi'(\theta_j)\,
   \Phi'(\theta_j)^{\top}(J\boldsymbol{m}) \notag\\
&= N\,(J\boldsymbol{m})^{\top}Q_N(\boldsymbol{m})\,J\,Q_N(\boldsymbol{m})\,(J\boldsymbol{m}).
\end{align}
Since $Q_N \to Q^*=O(1)$ and $J\boldsymbol{m}=O(\|\boldsymbol{m}\|)$, this gives
\begin{equation}
\mathrm{(II)} = O\bigl(N\|\boldsymbol{m}\|^2\bigr).
\end{equation}
Its contribution to the second term of~\eqref{eq:TrW_exact} is therefore
\begin{equation}
\frac{2\,|\mathrm{(II)}|}{\|\nabla H\|^4}
= \frac{O(N\|\boldsymbol{m}\|^2)}{O(N^2)}
= O\!\left(\frac{\|\boldsymbol{m}\|^2}{N}\right).
\end{equation}
This is suppressed by $N^{-1}$ relative to the collective quadratic correction
$O(\|\boldsymbol{m}\|^2)$ in $\mathrm{Tr}\,W_{\boldsymbol{\xi}}/N$.

\medskip
In conclusion, collecting the three blocks, the full Hessian-contraction term decomposes as
\begin{equation}
\frac{2\,\nabla H^{\top}(\mathrm{Hess}\,H)\nabla H}{\|\nabla H\|^4}
= \underbrace{O(N^{-1})}_{\text{kinetic}}
+ \underbrace{O\!\left(\|\boldsymbol{m}\|^3/N\right)}_{\text{diag.\ angular}}
+ \underbrace{O\!\left(\|\boldsymbol{m}\|^2/N\right)}_{\text{interaction}}.
\label{eq:hess_summary}
\end{equation}
In all three cases the contribution is suppressed relative to the terms
retained in $\mathrm{Tr}\,W_{\boldsymbol{\xi}}/N$: the background $1/c(\varepsilon)$
is $O(1)$, and the collective correction is $O(\|\boldsymbol{m}\|^2)$, both
without any $N^{-1}$ prefactor. The suppression mechanism is therefore
\emph{double}: every block is $O(N^{-1})$ in the thermodynamic limit,
and the dominant block~(II) is simultaneously $O(\|\boldsymbol{m}\|^2/N)$,
meaning it lies one order below the quadratic collective expansion.
The approximation
\begin{equation}
\frac{\mathrm{Tr}\,W_{\boldsymbol{\xi}}}{N}
= \frac{\Delta H/N}{\|\nabla H\|^2/N} + O(N^{-1})
\end{equation}
is therefore justified at leading order in $N$ and at any fixed order
in $\|\boldsymbol{m}\|$.

\printcredits

\end{document}